\documentclass[longbibliography,aps,pre,twocolumn,superscriptaddress,showpacs]{revtex4-1}

\usepackage{hyperref}
\usepackage{latexsym}
\usepackage{graphics}
\usepackage{textcomp}
\usepackage{amsmath}
\usepackage{comment}

\usepackage{xcolor,amssymb}
\definecolor{orange}{rgb}{1,0.5,0}

\begin{document}

\title{Simultaneous rheo-electric measurements of strongly conductive complex fluids}

\author{Ahmed Helal} 
\email[]{ahelal@mit.edu}
\affiliation{Massachusetts Institute of Technology, 77 Massachusetts Avenue, Cambridge MA 02139, USA}
\author{Thibaut Divoux}
\affiliation{Centre de Recherche Paul Pascal, CNRS UPR 8641, 115 avenue Schweitzer, 33600 Pessac, France}
\author{Gareth H. McKinley} 
\affiliation{Massachusetts Institute of Technology, 77 Massachusetts Avenue, Cambridge MA 02139, USA}

\date{\today}

\begin{abstract}
We introduce a novel apparatus designed for stress-controlled rheometers to perform simultaneous rheological and electrical measurements on strongly conductive complex fluids under shear. By means of a non-toxic liquid metal at room temperature, the electrical connection to the rotating shaft is completed with minimal additional mechanical friction, allowing for simultaneous stress measurements as low as 1~Pa. Motivated by applications such as flow batteries, we use the capabilities of this design to perform an extensive set of rheo-electric experiments on gels formulated from attractive carbon black particles, at concentrations ranging from 4 to 15\%~wt. First, experiments on gels at rest prepared with different shear history show a robust power-law scaling between the elastic modulus $G'_0$ and the conductivity $\sigma_0$ of the gels, i.e. $G'_0 \sim \sigma_0^\alpha$, with $\alpha =1.65 \pm 0.04$ independently of the gel concentration. Second, we report conductivity measurements performed simultaneously with creep experiments. Changes in conductivity in the early stage of the experiments, also known as the Andrade creep regime, reveal for the first time that plastic events take place in the bulk while the shear rate $\dot \gamma$ decreases as a weak power law of time. The subsequent evolution of the conductivity and shear rate allows us to propose a local yielding scenario that is in agreement with previous velocimetry measurements. Finally, to establish a set of benchmark data, we determine the constitutive rheological and electrical behavior of carbon black gels. Corrections first introduced for mechanical measurements regarding shear inhomogeneity and wall slip are carefully extended to electrical measurements to accurately distinguish between bulk and surface contributions to the conductivity. As an illustrative example, we examine the constitutive rheo-electric properties of five carbon black gels of different grades, and demonstrate the relevance of the novel rheo-electric apparatus as a versatile characterization tool for strongly conductive complex fluids and their applications.      
\end{abstract}

\pacs{83.80.Kn,47.57.Qk,77.84.Nh,81.05.U-}

\maketitle

\section{Introduction}
\label{intro}

From wormlike micellar fluids to colloidal glasses, complex fluids encompass a wide array of soft materials that all display a deformable and reconfigurable microstructure, which gives rise to intermediate mechanical properties between that of liquids and solids \cite{Balmforth2014,McKinley2015,Lu2013}. Under shear, the microstructure can rearrange and interact with the flow, which may in turn feed back on the flow itself triggering instabilities \cite{Coussot2007}. As such, complex fluids can exhibit a host of non-linear effects such as shear banding instabilities \cite{Fardin2012c,Divoux2016} as well as shear localization \cite{Ovarlez2009}, wall slip~\cite{Barnes1995,Kalyon2005,Buscall2010,Fardin2012b}, macroscopic fractures \cite{Persello1994,Pignon1996}, shear-induced structuration \cite{DeGroot1994,Montesi2004,Osuji2008,Negi2009,Grenard2010}, etc. 
The use of complementary local measurements such as X-ray scattering~\cite{Pujari2011} and Neutron spectroscopy~\cite{Eberle2012} to extract structural information on the one hand, or MRI~\cite{Bonn2008,Callaghan2008}, particle tracking \cite{Rajaram2010a}, dynamic light scattering and ultrasound imaging \cite{Manneville2008,Gallot2013} to extract velocity profiles on the other hand, has provided valuable insights about complex fluid rheology through time-resolved measurements of the microstructure evolution under shear. 

Another microstructural probe, less frequently employed, is measurement of the electrical properties of the fluid (e.g. conductivity $\sigma$, permittivity $\epsilon$), which coupled with standard rheological measurements also provides additional information on the evolution of the fluid microstructure under flow~\cite{Mewis1987, Genz1994, Capaccioli2007,Bauhofer2010,Alig2012,Amari1990,Crawshaw2006}.    
Previous rheo-electric studies have focused mostly on dielectric complex fluids such as polymeric materials~\cite{Watanabe1997, Watanabe1999a, Lou2013}, liquid crystals~\cite{Capaccioli2007}, colloidal suspensions~\cite{Crawshaw2006} and electrorheological fluids~\cite{Pan1998}, as well as weakly conductive materials ($\sigma\ll 1$~mS/cm) such as CNT nanocomposites~\cite{Alig2012, Alig2007, Alig2008, Bauhofer2010, Cipriano2008a, Schulz2010, Kharchenko2004} and weakly conductive carbon black gels~\cite{Watanabe2001,Genz1994, Mewis1987, Madec2014a, Youssry2013, Youssry2015a}. Typical commercial rheo-electric platforms involve the use of a rheometer coupled with an electrical measuring system such as an LCR meter that measures electric or dielectric constants simultaneously. In some studies, electrical measurements are performed using Alternating Current (AC) to extract the complex permittivity of the material~\cite{Capaccioli2007, Watanabe2001, Genz1994, Madec2014a} while other studies have focused on the use of Direct Current (DC) measurements to characterize the conductivity of weakly conductive materials under flow~\cite{Alig2012, Alig2008, Mewis1987, Schulz2010, Bauhofer2010}. However, rheo-electric studies of complex fluids are rare as they typically require the use of less common and expensive strain-controlled rheometers in which the rotary drive and stationary torque sensor are separate. Furthermore, most previous studies assume homogeneous flows and neglect non-linear kinematic phenomena discussed above such as slip and shear banding that regularly occur in conductive complex fluids \cite{Gibaud2010,Grenard2014}, and can lead to misinterpretation of rheological data \cite{Madec2014a, Youssry2013, Youssry2015a}.  

Characterizing the electrical properties of strongly conductive complex fluids ($\sigma\gg 1$ mS/cm) under flow is of practical relevance for numerous applications among which the large-scale manufacturing of conductive nanocomposites \cite{Alig2012,Schulz2010,Bauhofer2010,Liu2007} and the design of semi-solid flow battery electrodes \cite{Fan2014,Duduta2011,Li2013a,Youssry2013} are particularly noteworthy. For instance, the use of an embedded dispersed nanoconductor, such as carbon black, as an additive in semi-solid flow cells has been shown to increase the electrochemical activity throughout the volume of the cell, leading to higher rate performance and higher energy densities at lower cost \cite{Fan2014,Youssry2013,Wei2015}. As such, characterizing the evolution of the conductivity of battery slurries under imposed shear is crucial to optimize the design and operation of semi-solid flow cell \cite{Smith2014}. 

\begin{figure}
	\centering
	\resizebox{0.5\textwidth}{!}{\includegraphics{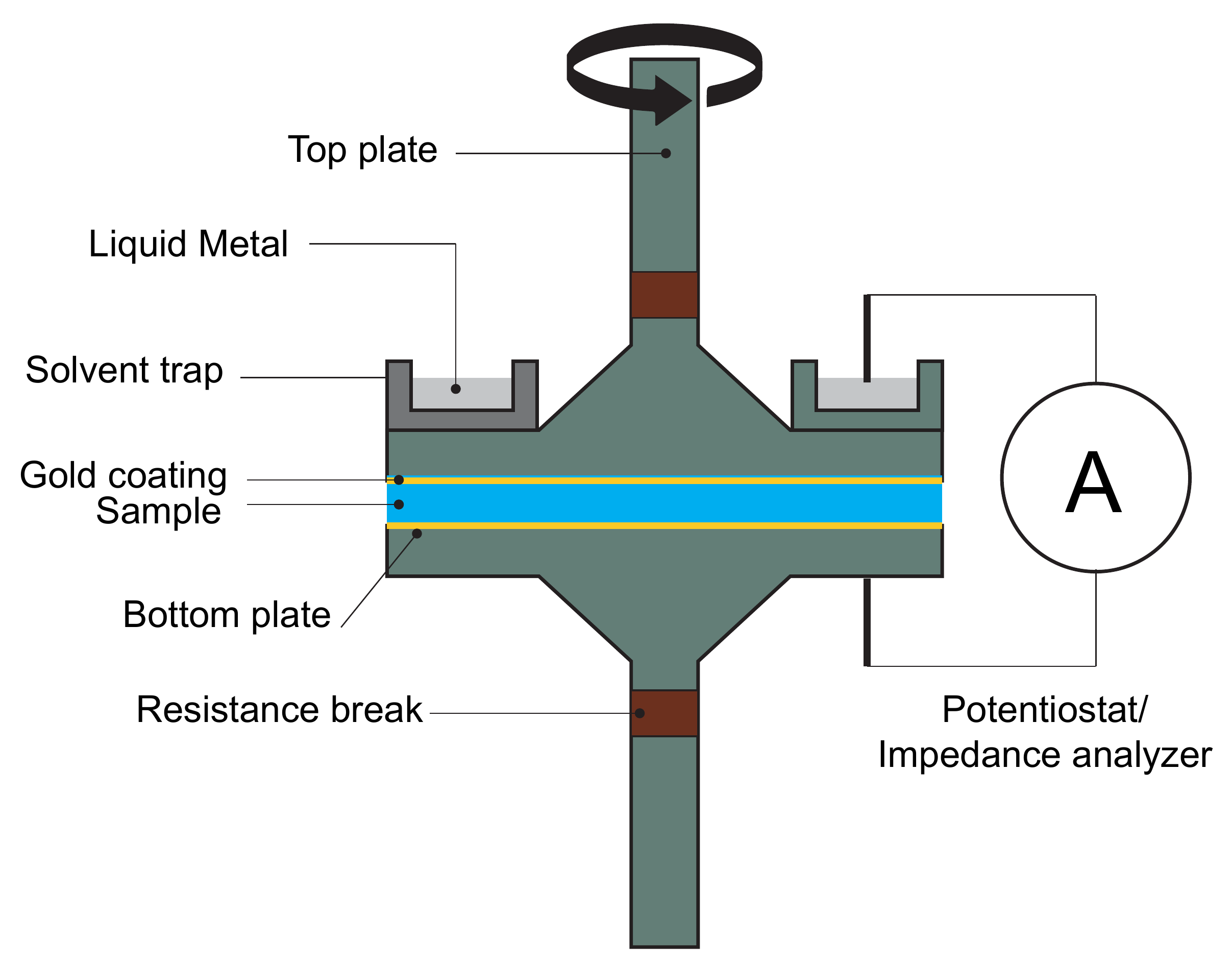}}
	\caption{Schematic of the rheo-electric test fixture. The sample is sandwiched between two parallel plates ($R=20$ mm) that form a two-electrode closed electrical circuit. The rotating upper plate is driven by a stress-controlled rheometer that applies stress and measures strain. Liquid metal (EGaIn) in the solvent trap provides a low-friction electrical contact to the rotating shaft. The plates are sputter-coated in gold to reduce contact resistance.}
	\label{fig:1}       
\end{figure}

In the present work, we use a new custom-made rheo-electric apparatus to investigate the effect of shear on the conductivity of strongly conductive complex fluids. The goal of this paper is twofold. First, in section~\ref{sec:1}, we describe a novel rheo-electric test fixture developed for stress-controlled rheometers, that is characterized by a low mechanical friction, continuous electrical contact under rotation and low contact resistance. Second, in section~\ref{sec:3}, we aim at illustrating the relevance of this apparatus on a class of strongly conductive complex fluids, namely carbon black gels. In section~\ref{sec:3.1}, we measure simultaneously the electrical and mechanical properties of carbon black gels at rest, which leads to a generic power-law scaling between elastic modulus and conductivity of the gels. We then tackle the stress-induced yielding scenario of these gels in section~\ref{sec:3.2}, revealing the existence of bulk rearrangements associated with minute macroscopic strains during the primary creep regime. Finally, section~\ref{sec:3.3} is devoted to steady-state flow and to the accurate determination of the electrical and mechanical constitutive behavior of carbon black gels. We show that the corrections first introduced for mechanical measurements extend to electrical measurements and enable accurate conductivity measurements under shear. Section~\ref{sec:4} serves as a discussion and includes a comparison of the rheo-electric properties of CB gels formulated from different grades of carbon black. This comparison illustrates how the structure of CB particles greatly affects the rheo-electric properties of the resulting gels, and serves as a guideline for applications requiring complex fluids with high conductivities.

\section{Rheo-electric test fixture}
\label{sec:1}

\subsection{Description of the test fixture}
\label{sec:2.2}

Rheo-electric platforms typically involve coupling an electrical measuring system to a conventional torsional rheometer that is used to measure the rheological properties of the materials under investigation. In previous work, a number of different two-electrode geometries have been used including cylindrical Couette~\cite{Amari1990,Genz1994, Mewis1987}, plate-plate~\cite{Youssry2015a,Capaccioli2007,Cipriano2008a, Madec2014a}, ring geometries \cite{Alig2007,Alig2008,Alig2012,Bauhofer2010} and pipe flow \cite{Crawshaw2006}. The difficulty inherent in performing simultaneous rheo-electric measurements lies in the additional friction caused by the electrical connection to the rotating electrode (typically through a slip ring), which can induce large errors in the measurement of the torque, when measured at the same electrode. Controlled-rate rheometers are therefore generally preferred for this class of measurements as they allow the user to measure torque on the stationary electrode, thus decoupling the electrical and rheological measurements. However, rate-controlled rheometers are more expensive and less common than stress-controlled rheometers and less well-suited for measuring yielding transitions that are common in highly filled conductive materials. These limitations motivate the development of a new apparatus. 

\begin{figure*}[t]
	\centering
	\resizebox{1\textwidth}{!}{\includegraphics{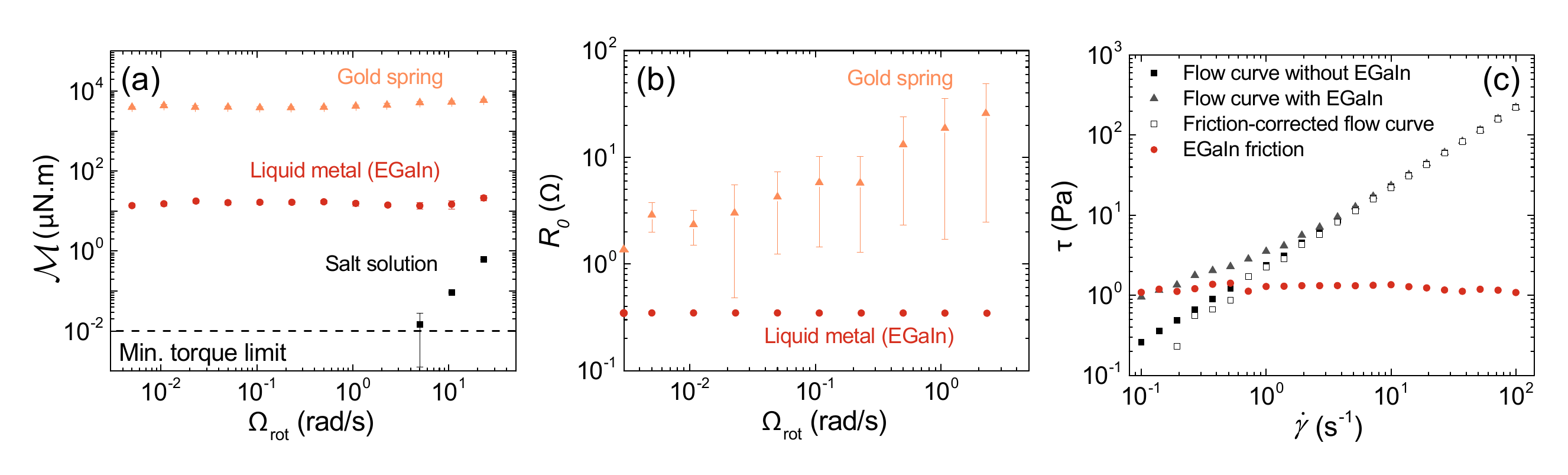}}      
	\caption{(a) Total measured torque $\mathcal{M}$ due to the friction of the electrical connection to the rotating shaft as a function of the angular velocity $\Omega_{rot}$ of the upper plate. Measurements are conducted in the absence of a sample and for different electrical contacts: a gold leaf spring, liquid metal (EGaIn) and a salt solution. (b) Setup DC resistance $\mathcal{R}_{0}$ measured in the absence of a sample as a function of $\Omega_{rot}$ for two electrical conductors: either a gold leaf spring or a liquid metal (EGaIn). Error bars are determined by calculating the standard deviation of the temporal signal over 180s. The use of a salt solution as a conductor does not allow for DC measurements as the ionic species will only polarize under DC current. (c) Measured flow curves showing shear stress $\tau$ vs shear rate $\dot{\gamma}$ for a Newtonian calibration oil and illustrating the friction correction due to EGaIn. Red filled circles represent the friction calibration performed with EGaIn in the absence of a sample prior to the flow curve measurement. Green triangles and black squares represent respectively the flow curves measured with and without the liquid metal. Open black squares represent the final corrected flow curve.}
	\label{fig:2}       
\end{figure*}

A schematic of the experimental test fixture, later referred to as the rheo-electric test fixture, and used to perform the simultaneous rheo-electric measurements is shown in Fig.~\ref{fig:1} [See also Fig.~S1 in the Supplemental Material for a picture]. The apparatus is built around two parallel plates (of radius $R = 20$~mm, and surface roughness $R_a = 0.1~\mu$m, as determined by a Mitutoyo surface profilometer) that form a two-electrode cell separated by a distance $H$ with the conductive sample sandwiched between them. The upper rotating plate is connected to a stress-controlled torsional rheometer (TA instruments, AR-G2) that applies a torque and measures the resulting shear strain. The plate-plate geometry is chosen here as it allows for a homogeneous electric field and can be used to correct for wall slip effects through measurements at multiple gap settings~\cite{Yoshimura1988}. However, the shear rate $\dot{\gamma}$ is not homogeneous in the parallel plate geometry and depends on the radial position $r$, as follows: $\dot{\gamma}(r)=r\Omega_{rot}/H$, where $\Omega_{rot}$ denotes the angular velocity of the upper plate. The impact of such inhomogeneity on the rheological measurements is taken into account and discussed in section~\ref{sec:3.3}. The geometry forms a closed electrical circuit that incorporates ($i$) the conductive sample, ($ii$) the plates, ($iii$) the electrical measuring system formed by a potentiostat (Solartron SI1287) with an Impedance Analyzer (SI1266) and ($iv$) the solvent trap (constructed from a machined annulus of SS316) filled with eutectic gallium-indium liquid metal (EGaIn, from GalliumSource). EGaIn is a liquid metal at room temperature which provides a safe non-hazardous alternative to other liquid metals such as mercury~\cite{Watanabe1999a} and is used to create a low-friction continuous electrical connection to the rotating shaft. Finally, to reduce the contact resistance, the plates are sputter-coated with gold using a Pelco SC-7 sputterer for 10~min. Unless otherwise noted, all rheo-electric tests were performed at a temperature $T = (23.0 \pm 0.3)^\circ$C, with a gap $H=750~\mu$m.

\begin{figure*}[t] 
	\centering
	\resizebox{1\textwidth}{!}{\includegraphics{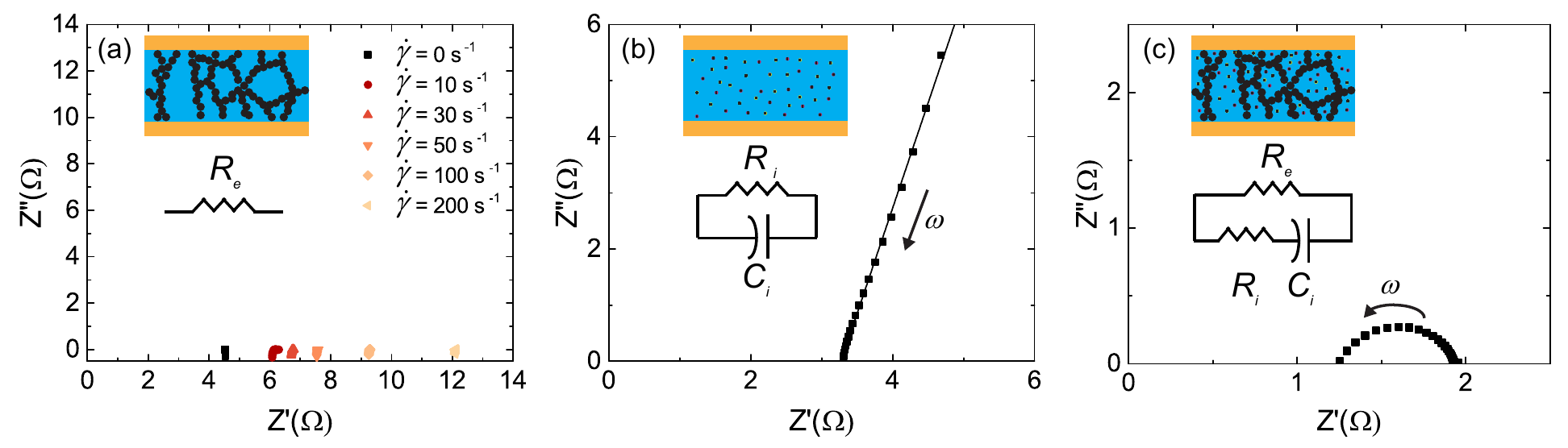}}     
	\caption{ Nyquist plots of complex impedance for three different classes of conductive complex fluids and their corresponding equivalent circuits: (a) Electronic conductor: suspension of attractive soot particles (carbon Black, VXC72R) dispersed in mineral oil at 8\%~wt.. The conductivity is measured at different shear rates coded in color: $\dot{\gamma}=0$, 10, 30, 50, 100 and 200~$\text{s}^{-1}$ from dark to light color; (b) Ionic conductor (Oakton 2764 $\mu$S/cm KCl solution) measured under static conditions for $H=750$ $\mu$m. The linear fit corresponds to the model blocking circuit shown in the inset. (c) Mixed conductor: Lithium polysulfide suspension containing 2.5M $\text{Li}_{2}\text{S}_{8}$, 0.5M LiTFSI and 1M $\text{LiNO}_{3}$ with 1.5vol\% Ketjenblack carbon measured under static conditions. AC tests were performed for $10^2\leq\omega\leq 10^4$~Hz.}
	\label{fig:3}       
\end{figure*}

Under static conditions, two types of electrical conductivity measurements are of interest: DC and AC tests. In a DC test, a constant potential $\phi_0=100$ mV is applied and the corresponding current $I$ is measured. The apparent DC conductivity $\sigma_{a}$ is then given by 
\begin{equation}
\sigma_{a}=\frac{H}{\pi R^2}\frac{I}{\phi_0} 
\label{eq1}
\end{equation}
where $H$ is the sample gap imposed by the rheometer. In an AC test, a sinusoidal potential $\phi=\phi_0 \sin(\omega t)$ is applied and the corresponding alternating current is measured. The complex impedance is defined as  
\begin{equation}
Z^*(\omega)=\frac{\phi}{I}=Z'-iZ''
\label{eq2}
\end{equation}
where $Z'$ and $Z''$ are, respectively, the real and imaginary contributions to the impedance. Note that $Z'$ and $Z''$ are the electrical analog of the real and imaginary contributions ($\eta'$ and $\eta''$ respectively) to the complex viscosity $\eta^*(\omega)$ in rheology, which will be of interest in section~\ref{sec:3} \cite{Gemant1935,Bird2012}. Finally, the complex impedance is related to the complex dielectric permittivity through the following expression \cite{Allen2001}:
\begin{equation}
\epsilon^*=\frac{H}{Z^*i\omega \epsilon_0 \pi R^2}
\label{eq3}
\end{equation}

In both AC and DC measurements, it is important to note that the measured resistances and impedances correspond to a sum of two contributions: a bulk resistance due to the sample resistance, and an interfacial contact resistance representing how readily the electrons can flow through the interface between the sample and the plates so that:
\begin{equation}
\mathcal{R}_{measured}\equiv\frac{\phi_0}{I}=\mathcal{R}_{contact}+\frac{H}{\sigma_{bulk}\pi R^2} 
\label{eq4}
\end{equation}
Traditionally, by conducting measurements at different gaps, the contribution of the contact resistance can be calculated in conjunction with Eq.~(\ref{eq4}) and corrected for~\cite{Allen2001}. For strongly conductive fluids, the relative contribution of the contact resistance becomes increasingly important as the bulk conductivity increases and cannot be neglected if accurate measurements are desired. Unless stated otherwise, the conductivity and impedance discussed hereafter correspond to the apparent quantity defined respectively in Eq.~(\ref{eq1}) and (\ref{eq2}). 

\subsection{Mechanical and electrical Calibration}
\label{sec:2.3}

We first discuss the mechanical calibration of the rheo-electric test fixture, and justify the preferential use of liquid metal in the electrical circuit compared to two other techniques, namely a gold spring and a salt solution that are commonly used in other existing rheo-electric configurations. On stress-controlled rheometers, friction acting on the rotating shaft during electrical measurements under shear can impede simultaneous rheological measurements. To quantify such friction, we have performed measurements of the torque $\mathcal{M}$ associated with the friction inherent to different electrical couplings: EGaIn, a gold leaf spring, and a salt solution. Results are reported as a function of the angular velocity $\Omega_{rot}$ of the upper plate in Fig.~\ref{fig:2}(a). The friction due to the liquid metal lies two orders of magnitude below that of the gold spring, and is constant over four decades of angular velocity, thus leading to a straightforward correction for mechanical friction. This constant friction arises primarily from the thin oxide layer that forms on top of the liquid metal when in contact with air \cite{Dickey2008}. Furthermore, the temporal fluctuations of the friction contribution to the stress can be neglected during simultaneous measurements in general, when operating at stresses larger than the background friction level (see Fig.~S2 in the Supplemental Materials).

Regarding the electrical calibration, the DC resistance $\mathcal{R}_{0}$ of the test fixture in the absence of any sample is measured as a function of angular velocity $\Omega_{rot}$ for a closed electrical circuit using either EGaIn or a gold spring [Fig.~\ref{fig:2}(b)]. The liquid metal leads to a low resistance $\mathcal{R}_{0}=0.3~\Omega$ that is independent of angular velocity, which makes the contribution to the electrical resistance from the test cell easy to take into account. On the contrary, the configuration involving the gold leaf spring shows a larger and much noisier resistance that increases with the plate rotation speed. We also emphasize that the use of a salt solution only allows for AC measurements at low potentials ($\phi_0\leq 10$~V) and tends to have a complex electrical signature due to ion mobility that can complicate electrical corrections \cite{Allen2001}. 

From both a mechanical and electrical standpoint, the test configuration using EGaIn is the most suitable for rheo-electric measurements on stress-controlled rheometers. As a result of the low friction, we are able to perform simultaneous measurements of rheology and conductivity, an example of which is shown in Fig.~\ref{fig:2}(c) on a calibration oil (N1000, Cannon Instrument Company). The additional friction due to the EGaIn is constant at about 1.2~Pa. By subtracting the latter value from the flow curve measured using the same experimental fixture, one recovers with high accuracy the true flow curve of the calibration oil measured in the absence of the electrical connections, at least for shear stresses $\tau \geq  1.2$~Pa. For shear stresses $\tau< 1.2$~Pa, the corrected data departs appreciably from the constitutive equation showing that although conductivity measurements may still be performed, stress measurements should be conducted separately in this low stress range, i.e. without the upper electrode immersed in the liquid metal (EGaIn). 

\subsection{Benchmark electrical measurements}
\label{sec:2.4}

To illustrate the ability of the EGaIn rheo-electric fixture to perform accurate conductivity measurements, we conducted a series of benchmark AC measurements ($\phi_0=100$~mV, $10^2\leq\omega\leq 10^4$~Hz) on different classes of conductive complex fluids, namely electronic, ionic and mixed conductors. The AC response of these fluids are reported in Fig.~\ref{fig:3}, where the imaginary part $Z''$ of the complex impedance is plotted vs the real part $Z'$ of the complex impedance. Such a representation, also known as a Nyquist plot \cite{Allen2001}, is analogous to a Cole-Cole plot for rheological measurements. First, we report the response of an electronic conductor, here a suspension of attractive soot particles (carbon black VXC72R, Cabot) dispersed in mineral oil at $8$\%~ wt. that is measured under steady simple shear flow. The electrical resistance of this suspension is well represented by a pure ohmic resistor and corresponds to a series of points that sit on the horizontal axis of the Nyquist plot [Fig.~\ref{fig:3}(a)]. As the shear rate is increased, the resistance increases approximately three-fold due to the disruption of the floculated carbon black network under shear. Second, we report an example of an ionic conductor (Oakton 2764 $\mu$S/cm KCl solution, Cole-Parmer) measured under static conditions for $H=750$~$\mu$m and at $T=25.0^{\circ}$C [Fig.~\ref{fig:3}(b)]. This suspension is represented by an ohmic resistor in parallel with an imperfect capacitor [$Z^*(\omega)=\mathcal{R}_i+1/(i\omega C_i)^{\alpha}$], represented by a constant phase element of phase $\alpha$. In the Nyquist plot, it appears as a line of slope $\tan(\alpha\pi/2)$ that intercepts with the horizontal axis at $Z'=\mathcal{R}_i$, which represents the ionic resistance of the solution. The experimental data are indeed well described by a blocking circuit [black fit in Fig.~\ref{fig:3}(b)]. Supplemental measurements at various gaps and Eq.~(\ref{eq4}) allow us to correct for the contact resistance (see Fig.~S3 in the Supplemental Materials) and thus infer the bulk conductivity of the ionic solution. The contact resistance and bulk conductivity are found to be respectively $\mathcal{R}_c=0.94$~$\Omega$ and $\sigma_{b}=(2.76 \pm 0.01)$~mS/cm at $T=25.0^\circ$C, in excellent agreement with the value provided by the manufacturer ($\sigma=2.764$ mS/cm). Finally, we consider a third sample which is an example of a semi-solid battery material that consists of a mixed conductor (Lithium polysulfide suspension containing 2.5M $\text{Li}_{2}\text{S}_{8}$, 0.5M LiTFSI and 1M $\text{LiNO}_{3}$ with 1.5vol\% Ketjenblack EC600JD carbon black). Its equivalent circuit, which is the mechanical analog of the Jeffreys model in rheology \cite{Bird1987}, corresponds to a depressed semi-circle that intersects the horizontal axis at $Z'=\mathcal{R}_e\mathcal{R}_i/(\mathcal{R}_e+\mathcal{R}_i)$ and $Z'=\mathcal{R}_e$ respectively. The Nyquist plot of this slurry, determined under static conditions using the rheo-electric fixture is shown in Fig.~\ref{fig:3}(c). The ionic and electronic conductivities are estimated to be $\sigma_i=2.6$ mS/cm and $\sigma_e=3.7$ mS/cm respectively, in agreement with measurements previously reported on the same material~\cite{Fan2014}.  

To conclude, the use of gold-sputtered plates and EGaIn to close the electrical circuit allow for accurate electrical measurements on a wide variety of strongly conductive complex fluids, which demonstrate the versatility of the EGaIn rheo-electric test fixture. In the rest of the paper, we employ the rheo-electric fixture to investigate in detail one of the afore-mentioned model systems, namely carbon black gels.

\section{Rheo-electric study of carbon black gels}
\label{sec:3}

We focus on carbon black dispersions as a case study of a strongly conductive complex fluid to illustrate the additional insight that can be gained from this rheo-electric apparatus. Carbon black (CB) which refers generically to colloidal soot particles produced from the incomplete combustion of fossil fuels, is utilized extensively in a wide variety of industrial applications including batteries, paints, coatings as well as filled rubbers and tires. CB serves as an additive to convey protection from ultra-violet light, increase toughness, improve processability and to enhance electrical conductivity \cite{Donnet1993,Yearsley2012}. Composed of permanently fused ``primary" particles whose diameter is about 30~nm \cite{Samson1987}, CB particles are of typical size 200-500~nm and display short range attractive interactions of typical strength 30~$k_BT$ \cite{Trappe2007}. When dispersed in a liquid hydrocarbon, carbon black particles form a space-spanning network, even at small volume fractions (typically 1 to 5\%) \cite{VanderWaarden1950,Trappe2000}. The resulting gels are reversible, as they are destroyed by large shear and reform once the flow is stopped. The elastic modulus and yield stress depend on the flow history which can be used to tune the solid-like properties of the gel at rest \cite{Osuji2008,Ovarlez2013}. Finally, shear-induced changes in microstructure are known to strongly affect the electrical properties of carbon black gels as the percolated network ruptures under shear \cite{Mewis1987,Amari1990,Liu2007}. Depending on the choice of the carbon black, the conductivity of the gel either decreases by several orders of magnitude as the material is sheared and undergoes a conductor to insulator transition \cite{Amari1990}, or, in the case of more conductive samples, the conductivity only decreases by a small factor under shear \cite{Mewis1987}. 
 
In this work, a number of CB dispersions are prepared in the absence of any additional dispersant by mixing CB particles of different grades [VXC72R, Monarch 120 and Elftex 8 from Cabot chemicals (Billerica, MA), and KetjenBlack EC600 from Akzo Nobel (USA)] into a light mineral oil (Sigma, density 0.838 g/$\text{cm}^3$, viscosity 20 mPa.s) as described in ref.~\cite{Gibaud2010} at weight concentrations $c$ ranging from 4 to 15\%~wt. The samples are sonicated for 4~hours and mixed thoroughly prior to any use. In the following sections, we mainly report data obtained with a 8\%~wt. carbon black gel (VXC72R) which has been extensively studied in the literature \cite{Osuji2008,Grenard2010,Sprakel2011,Grenard2014}. Such a gel is used here as a model conductive complex fluids as it shows neither chemical aging nor any evaporation, and allows for reproducible measurements over long durations. Furthermore, the VXC72R carbon black has been specifically developed for high conductivity applications and is an ideal system to explore the rheo-electric properties of strongly conductive complex fluids. Using the rheo-electric test fixture characterized in section~\ref{sec:1}, we first determine the electrical and mechanical solid-like properties of the CB gel at rest, which is brought to rest through different flow cessation protocols (section~\ref{sec:3.1}). We then focus on transient flows and perform creep experiments to study the yielding behavior of the CB gel from an electrical perspective (section~\ref{sec:3.2}). Finally, we report an extensive study of steady shear flows of the CB gel, and show how electrical measurements allow us to detect flow heterogeneities and wall slip (section~\ref{sec:3.3}). To conclude, a discussion section including a quantitative comparison of the rheo-electric properties of CB gels, prepared with different grades of CB, shows how the nature of the CB particles greatly affects the ensuing rheo-electric properties of the gels. Such analysis, made possible with the rheo-electric fixture introduced here, is key for numerous applications where the selection of an appropriate conductive sample should obey specific design constraints on both the mechanical and electrical properties (section~\ref{sec:4}). 

\subsection{Tuning rheo-electric properties of carbon black gels using shear history}
\label{sec:3.1}

\begin{figure}
	\resizebox{0.5\textwidth}{!}{\includegraphics{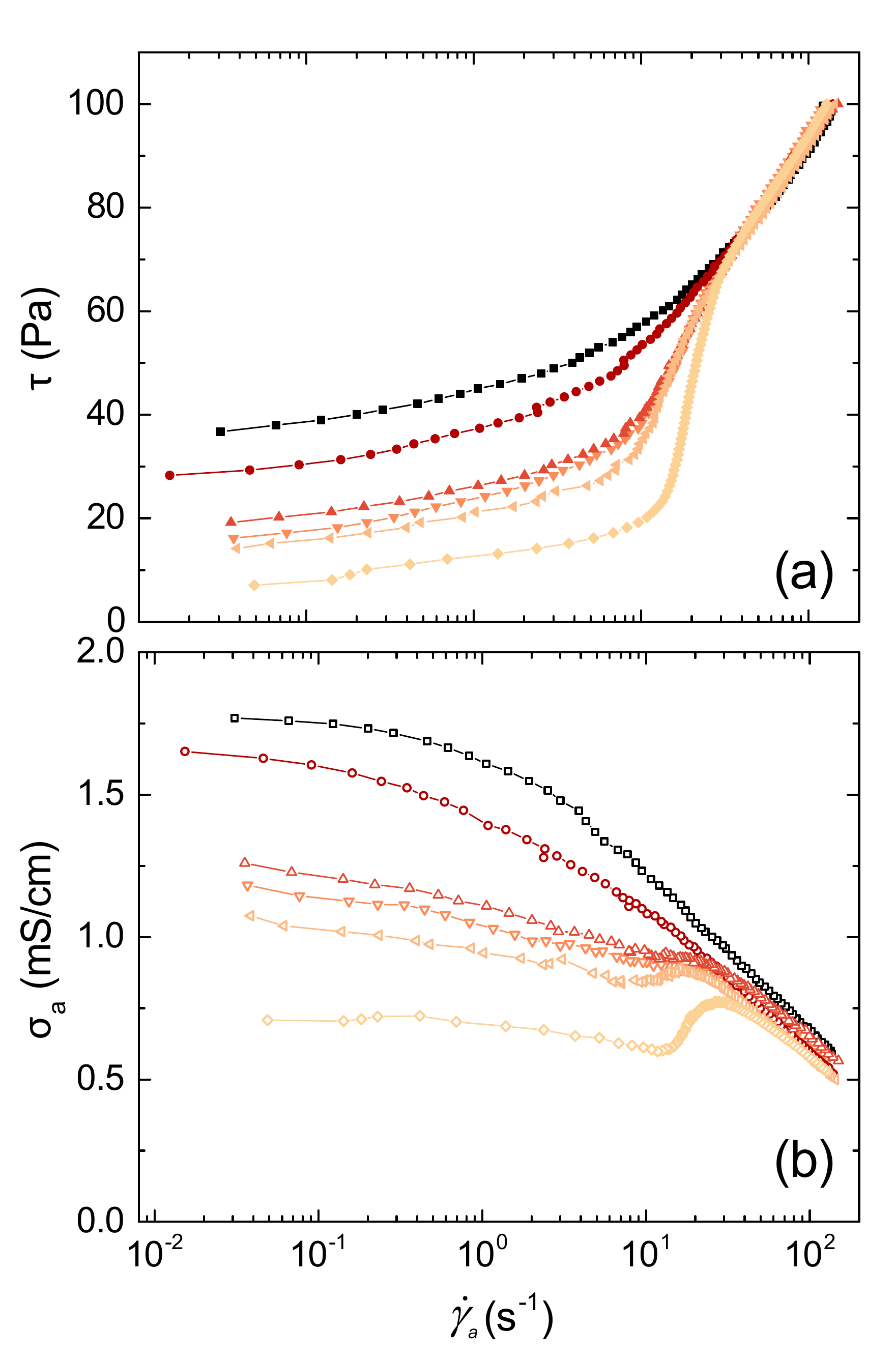}}
	\caption{(a) Stress $\tau$ vs apparent shear rate $\dot{\gamma}_{a}$ (b) Apparent conductivitiy $\sigma_{a}$ vs the apparent shear rate $\dot{\gamma}_a$ measured in a 8\%~wt. VXC72R carbon black gel during decreasing ramps of stress from $\tau=100$ Pa to $\tau=0$ Pa. Each test was performed in $N=100$ steps of 1 Pa each and different duration per point $\Delta t$ was used for each experiment. From top to bottom: $\Delta t=1,6,15,21,30 \text{ and } 120$ s.}
	\label{fig:4}       
\end{figure}

\begin{figure}[t]
	\centering
	\resizebox{0.5\textwidth}{!}{\includegraphics{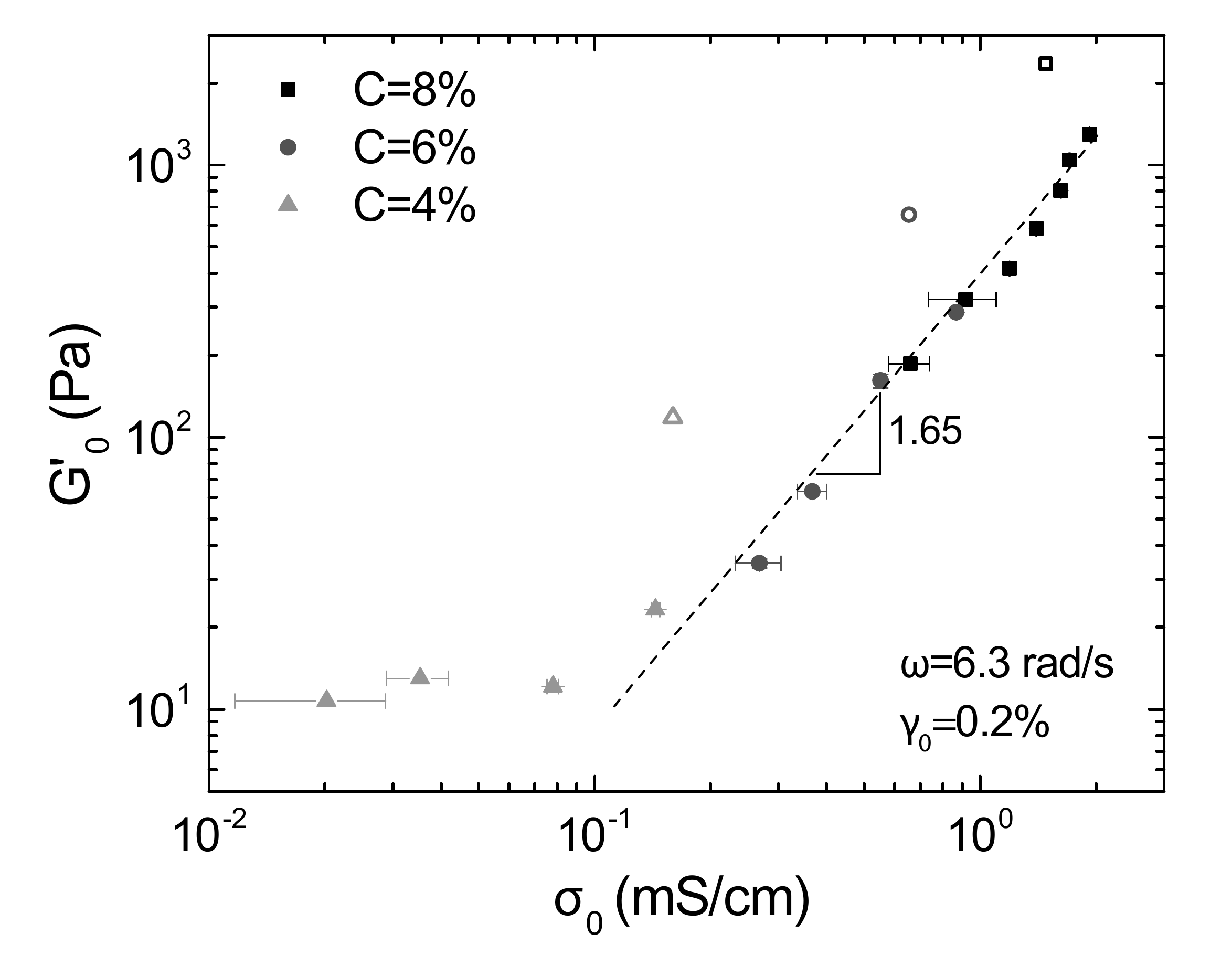}}
	\caption{Elastic modulus $G'_0$ of carbon black gels vs conductivity $\sigma_0$ measured at rest, 90~min after the end of a decreasing stress ramp from 100~Pa to 0~Pa and composed of 100 steps. Each point corresponds to different step durations $\Delta t$ during the ramp, from $\Delta t=1$~s to 120~s, while symbols stand for different concentrations of carbon black (VXC72R): $c=$4\%~wt. ($\blacktriangle$), 6\%~wt. ($\bullet$) and 8\%~wt. ($\blacksquare$). Open symbols stand for an instantaneous quench from $\tau=100$ Pa to $\tau=0$~Pa. Data for the 8\%~wt. gel corresponds to the experiment reported in Fig.~\ref{fig:4}. Oscillatory measurements were performed at a strain $\gamma_0=0.2$\% and a frequency $f=1$~Hz. The dashed line corresponds to the best power-law fit of the data for $\sigma_0 \geq 0.1$~mS/cm, i.e. $G'_0=\tilde{G'_0}\sigma_0^{\alpha}$, with $\tilde{G'_0}=$400~Pa.(cm/mS)$^{\alpha}$ and $\alpha=1.65\pm0.04$.} 
	\label{fig:5}       
\end{figure}

\begin{figure*}[t!]	
	\resizebox{1.02\textwidth}{!}{\includegraphics{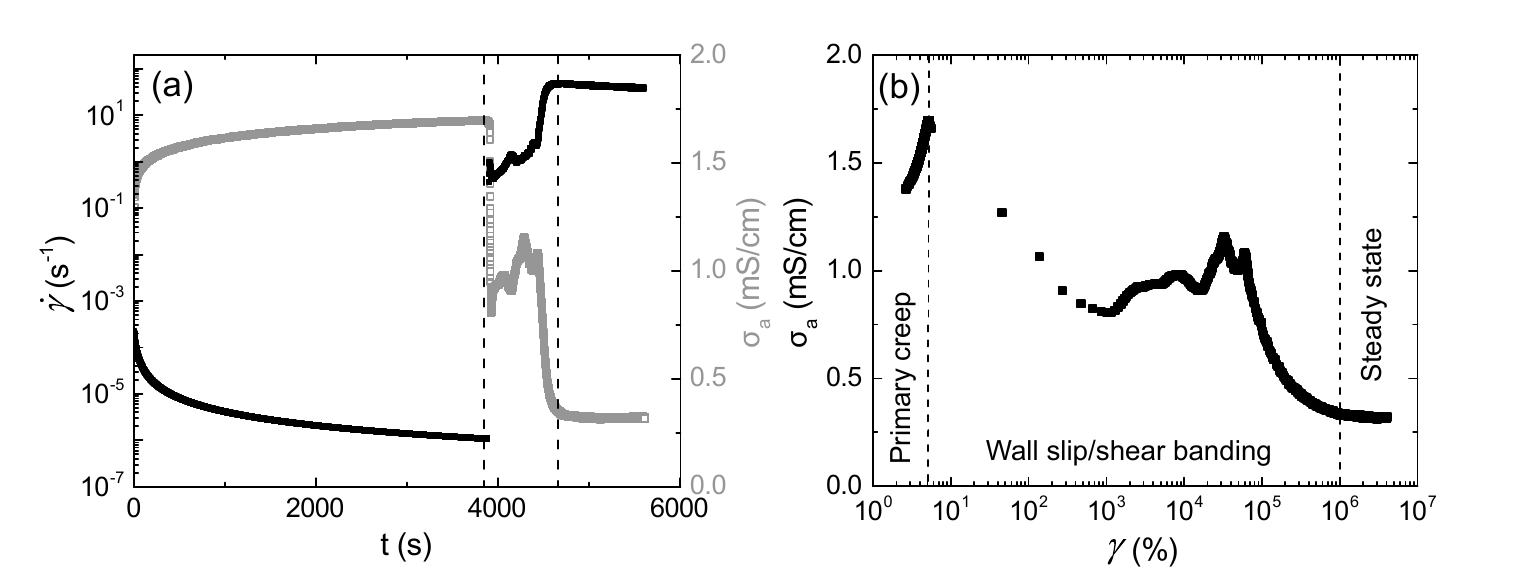}}
	\caption{Creep experiment in a 8\%~wt. carbon black gel (VXC72R) at $\tau=22$ Pa. (a) Shear rate response $\dot{\gamma}(t)$ and apparent conductivity response $\sigma_{a}(t)$. The shear rate was derived numerically from the strain measured by the rheometer to increase the precision. The jump observed in the curves at $t=3800$~s marks the onset of a total wall slip regime followed by the growth of a transient shear band \cite{Grenard2014} before the sample reaches a homogeneous steady state. (b) Apparent conductivity $\sigma_{a}$ vs strain $\gamma$ for the same data set. In both (a) and (b) the vertical dashed line marks the three different flow regimes discussed in the text: primary creep, heterogeneous flow and steady-state flow.}
	\label{fig:6}       
\end{figure*}

Recent studies have shown that the mechanical properties of weak attractive gels can be tuned by controlling their shear history \cite{Ovarlez2013, Koumakis2015,Yearsley2012}: abrupt flow cessation or ``quenching" leads to highly elastic gels, whereas slow cessation of shear leads to weak solids characterized by a vanishingly small yield stress for decreasing flow cessation rate. In this subsection, we investigate the evolution of the electrical properties concomitantly to the elastic properties of the gel prepared by flow cessation experiments at various cessation rates. Ramps of decreasing stress from $\tau=100$~Pa to $\tau=0$~Pa bringing the sample at complete rest are performed in $N=100$ successive steps. The upper stress value is applied for 5~min before the start of each decreasing ramp, and is chosen to be large enough to ensure the sample is fully fluidized and that any previous flow history is erased. The flow cessation rate is controlled through the duration of each step $\Delta t$, that is varied in each test from 1~s to 120~s. The resulting flow and conductivity curves are plotted in Fig.~\ref{fig:4}. As the stress is ramped down, the shear rate decreases and the conductivity concomitantly increases. Such results coincide with the fact that the gel transitions from a fully fluidized state at high shear rates to a solid-like state at rest. The gels structure builds up, forming a percolated network which accounts for the solid-like behavior and the conductivity increase. For $\dot{\gamma}_a> 30$~s$^{-1}$, the flow and conductivity curves obtained at different cessation rates collapse, indicating that in this region the gel is fully fluidized, and not sensitive to the flow history. At lower shear rates, both the flow and conductivity curves become strongly sensitive to the flow cessation rate: slower ramps (or increasing shearing durations $\Delta t$ spent at each rate) lead to lower yield stress (defined as the stress extrapolation for vanishing shear rates) and lower apparent conductivity. These results are associated with the growth of a solid-like structure as the gel comes to a stop, whose formation kinetics are strongly affected by $\Delta t$. 

To characterize the final state of the gel once the stress ramp is over, we record both the terminal value of the conductivity $\sigma_0$, and the linear elastic moduli $G'_0$ determined by oscillations of small amplitude ($\gamma=0.2$\%) at $f=1$~Hz. These measurements are performed 90~min after the end of the stress ramp to ensure that mechanical and electrical equilibrium is reached [see Fig.~S4 in the Supplemental Material]. Repeating the experiment for various step durations $\Delta t$ allows us to extract a series of pairs \{$G'_0$, $\sigma_0$\} that are plotted in Fig.~\ref{fig:5}.  
The elastic modulus $G^{'}_0$ displays a power-law increase with the apparent conductivity at rest $(\sigma_0)$ showing that there is a correlation between the mechanical and electrical properties of the gel at rest: slower ramps lead to softer and less conductive gels, a signature of lower connectivity in the sample-spanning network of CB aggregates. The latter results hold true at lower concentrations of carbon black particles. Data obtained following the same protocol with 4\%~wt. and 6\%~wt. carbon black gels [respectively ($\blacktriangle$) and ($\bullet$) in Fig.~\ref{fig:5}] show the same result and collapse on a master curve for $\sigma_0 \geq 0.1$~mS/cm, where $G'_0 \propto \sigma_0^{\alpha}$, with $\alpha=1.65\pm0.04$. The latter result hints at a unique gelation scenario kinetically driven by shear and not sensitive to the particle concentration, at least for strong enough gels with $G'_0 \gtrsim 10$~Pa. 

Finally, we have also performed sudden quenches, that is the CB gel is brought from $\tau=100$ Pa to $\tau=0$ Pa within 1~s. In that case, the conductivity is comparable to that obtained with the fastest ramp previously explored (of total duration 100~s), but the elastic modulus is much larger so that the rheo-electric properties at rest [open symbols in Fig.~\ref{fig:5}] do not fall onto the same master curve as the ramp tests of longer duration. However, we emphasize that the terminal elastic modulus still appears to grow as a power-law function of the conductivity, and the exponent is compatible (within error bars) with the value determined for softer flow cessations. For the three concentrations examined, abruptly stopping the shear flow leads to a gel of enhanced elasticity which hints at a specific gel microstructure that is well adapted to sustain shear and yet not more conductive (hence percolated) than gels formed in a fast ramp of 100~s. These results show that the electrical properties of carbon black gels can be tuned through the shear history in parallel to the mechanical properties.

\subsection{Stress-induced yielding}
\label{sec:3.2}

In this section, we turn to the transient behavior of carbon black gels and investigate the evolution of the rheo-electric properties of the 8\%~wt. CB gel (VXC72R) during its stress-induced yielding transition. Prior to each creep experiment, the CB gel is presheared at $\dot{\gamma}=200$~s$^{-1}$ for 5~min to erase previous flow history, and left to rebuild for 20~min at $\tau=0$~Pa. A constant stress is then applied from $t=0$ and the resulting shear-rate response $ \dot{\gamma}(t)$ and apparent conductivity response $\sigma_a(t)$ are monitored simultaneously. Results for different applied shear stresses are reported in Fig.~S5 of the Supplemental Material.

\begin{figure}[t!]
	\centering
	\resizebox{0.47\textwidth}{!}{\includegraphics{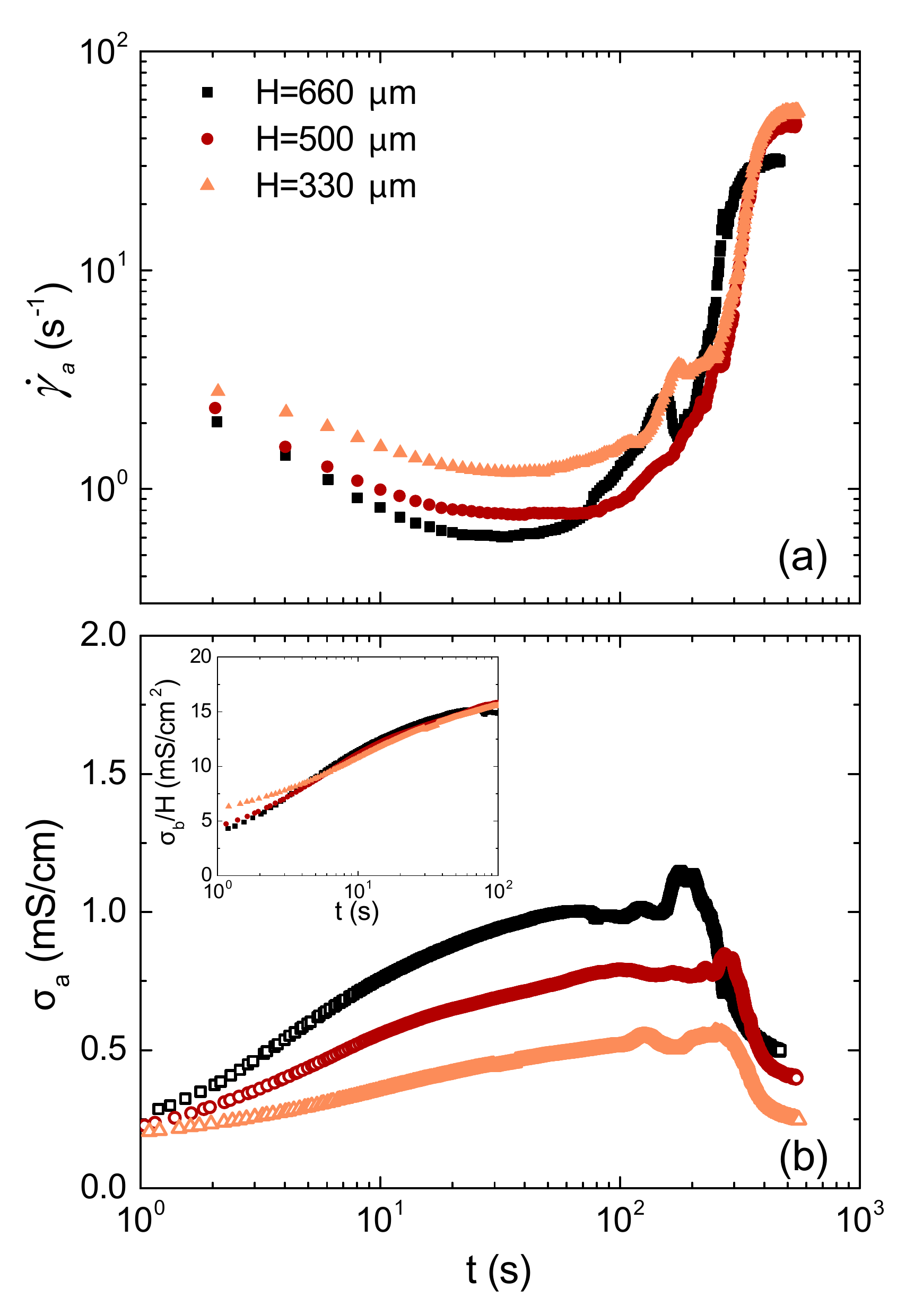}}
	\caption{Creep experiment in a carbon black VXC72R suspension ($c=8\%$ wt.) at $\tau=30$ Pa for different gaps ($H= 660, 500 \text{ and } 330$ $\mu$m). (a) Shear rate response $\dot{\gamma}(t)$ (b) Apparent conductivity response $\sigma_{a}(t)$. The inset shows the estimated bulk conductivity $\sigma_b(t)$ normalized by the gap $H$. The contribution of the contact resistance was estimated using Eq.~(\ref{eq4}) and the measurements at multiple gaps for $t=0$~s.} 
	\label{fig:7}       
\end{figure}

All the creep experiments display three characteristic regimes that we illustrate in Fig.~\ref{fig:6}(a) for the case of an experiment conducted at $\tau=22$~Pa. ($i$) First, for $t<3800$~s the shear rate decreases as a power law: $\dot \gamma(t) \sim t^{-0.9}$ while the apparent conductivity increases as a weak power law: $\sigma_a(t) \sim t^{0.035}$ [See Fig.~S6 in the Supplemental Material for a logarithmic plot of the same data]. ($ii$) Second, for $3800$~s~$\leq t \leq 4600$~s, a jump in the shear rate is observed and the apparent conductivity drops concomitantly by half and fluctuates. ($iii$) Third, for $t>4600$~s, the shear rate undergoes a second steep increase while the apparent conductivity drops, before both observables tend towards a final steady-state value. The mechanical response alone of this ``delayed yielding" has been previously reported in a Couette geometry \cite{Gibaud2010,Grenard2014}. Coupled with time-resolved velocimetry, the latter studies have shown that the three regimes can be interpreted as a succession of ($i$) homogeneous deformation above the micron scale followed by ($ii$) a total wall slip regime associated with the first rapid increase of $\dot \gamma (t)$, and the growth of a transient shear band [second increase of $\dot\gamma(t)$] to finally reach ($iii$) a homogeneous flow.   

The unique signature of these regimes also appears in the apparent conductivity response which provides further insights into the yielding behavior. Indeed, in the first regime, the increase in apparent conductivity clearly demonstrates that either the bulk microstructure of the sample, or the number of contacts between the sample and the wall, is evolving. The latter result highlights the high sensitivity of the rheo-electric test fixture in the first regime where the sample deformation remains extremely low [$\gamma \lesssim 4$\% for $t<$3800 s] and where other techniques, including ultrasonic velocimetry  \cite{Grenard2014}, fail to provide time-resolved data due to a lack of resolution. In the intermediate regime, the decrease in the apparent conductivity is compatible with the onset of slip which is expected to increase the electrical resistance at the interface. Moreover, the oscillations observed in the apparent conductivity are strongly reminiscent of the stick-slip like motion of the carbon black gel reported in ref. \cite{Grenard2014}, where alternating plug and shear flow regions have been observed. Finally, in the third regime, the conductivity reaches its lower, constant value -- which is in agreement with a fully fluidized sample, that makes uniform electrical contact with the wall, flowing homogeneously in steady state . 

A cross-plot of the apparent conductivity $\sigma_a$ vs strain $\gamma$ is shown in Fig.~\ref{fig:6}(b). In this representation, the delayed yielding and the three regimes discussed above are readily observed. In the primary creep regime, we observe a linear increase of apparent conductivity with strain, a signature of the increase of connectivity within the carbon black network in bulk or at the interface. In the intermediate regime, wall slip and shear banding result in the sudden decrease of apparent conductivity with strain as the sample fails at the interface, increasing the contact resistance. Finally, in the third regime, the apparent conductivity reaches a constant value as the strain increases, an indication that the system is fully fluidized and has reached steady state. 

Finally, to determine whether the increase of the conductivity during the primary creep regime is associated with bulk rearrangements, or with the evolution of the sample in the vicinity of the plates, we have performed supplemental creep experiments at $\tau=30$~Pa for gaps of different height $H=660,$ 500 and 330~$\mu$m. The corresponding shear rate and apparent conductivity responses are shown in Fig.~\ref{fig:7}. For the three gaps studied, the shear-rate responses display a similar evolution indicating that the material is subjected to the same delayed yielding scenario in time [Fig.~\ref{fig:7}(a)]. However, the various apparent conductivity responses exhibit significant differences: in the early stage of the experiment, all the apparent conductivity responses start at the same value, but the conductivity measured for larger gaps increases more rapidly [Fig.~\ref{fig:7}(b)]. To determine the bulk contribution, we estimate the contact resistance by extrapolating the apparent conductivity in the limit $t=0$~s. Then, using Eq.~(\ref{eq4}), we compute the bulk conductivity $\sigma_b$ for each experiment. This is then further normalized by the gap height $H$ [inset in Fig.~\ref{fig:7}(b)]. In the entire primary creep regime ($t \lesssim 50$~s), we observe that $\sigma_b/H$ is independent of the gap height, which demonstrates that the increase in apparent conductivity during the primary creep regime is due to a change in bulk of the sample properties. The deformation of the sample microstructure under external stress appears sufficient to trigger the pairing of attractive CB particles leading to an enhanced connectivity and thus a larger conductivity. In that framework, the faster increase of the conductivity at a larger gap can be interpreted as an increasing rate of plastic rearrangements for larger gaps. 

To conclude this section, simultaneous time-resolved rheo-electric measurements provide a more detailed picture of the yielding scenario which is in general agreement with previous velocimetry measurements \cite{Gibaud2010,Grenard2014}. As a key novel result, time-resolved measurements of the conductivity during the primary creep regime allow us to reveal the existence of stress-induced bulk rearrangements in the sample microstructure, which correspond to very small values of the macroscopic imposed strain.\\

\subsection{Steady-state measurements}
\label{sec:3.3} 
 
\begin{figure}
	\centering
	\resizebox{0.47\textwidth}{!}{\includegraphics{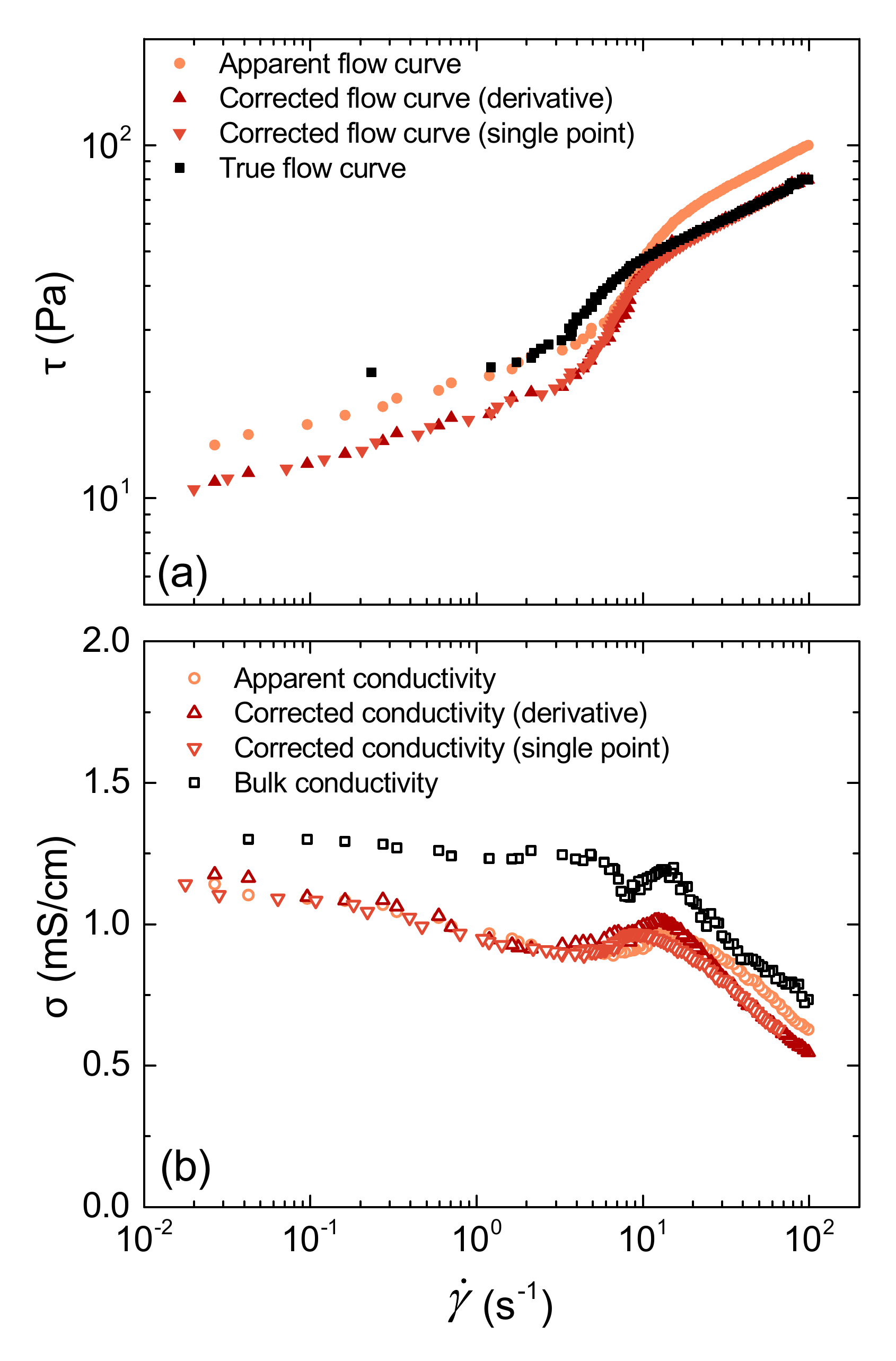}}	
	\caption{Steady-state flow and conductivity curves in a 8\%~wt. carbon black gel (VXC72R). (a) Flow curves showing the effects of the shear inhomogeneity correction (derivative and single point methods) as well as the slip correction. The ``corrected'' flow curves are obtained after correcting for shear inhomogeneity alone, while the true flow curve is obtained after correcting for both shear inhomogeneity and slip effects. (b) Conductivity curves showing the effects of the shear inhomogeneity correction as well as the contact resistance correction. The ``corrected'' conductivity curves are obtained after correcting for shear inhomogeneity alone, while the bulk conductivity curve is obtained after correcting for both shear inhomogeneity and contact resistance effects.}
	\label{fig:8}       
\end{figure}

After studying the transient rheo-electric behavior of CB gels, we now focus on steady-state shear measurements. The determination of the true constitutive equation is crucial for numerous applications, where the true bulk behavior must be accurately modeled \cite{Alig2012,Schulz2010,Bauhofer2010, Fan2014,Duduta2011,Li2013a,Youssry2013}, whereas numerous studies in the literature report only approximative or apparent flow curves \cite{Madec2014a, Youssry2013, Youssry2015a}. In this last section, we take advantage of the rheo-electric test fixture to characterize accurately the constitutive rheological behavior together with the conductivity of CB gels under flow by taking into account the inhomogeneity inherent to the parallel plate geometry, the slip of the sample at the wall and the contact resistance. 
A stress ramp of decreasing values from $\tau=$100~Pa to 0~Pa and composed of $N=100$ steps, each of duration $\Delta t=30$~s is performed on a 8\%~wt. CB gel (VXC72R). The choice of ramping the stress downwards and at a slow enough pace is made to insure that the rheological and electrical properties are close to steady state at all shear rates explored. The resulting flow and conductivity curves, as well as the effects of the different corrections detailed below, are shown in Fig.~\ref{fig:8}(a).

We first discuss the mechanical corrections. The shear rate $\dot{\gamma}$ is not homogeneous across the parallel plate, but depends on the radial position $r$ as $\dot{\gamma}(r)=r\Omega_{rot}/H$, where $\Omega_{rot}$ is the angular velocity of the upper plate. The corrected flow curve can be obtained using the apparent shear rate at the rim $\dot{\gamma}_{a,R}=R\Omega_{rot}/H$ and by computing the stress at the rim $\tau_R$ by means of the following expression \cite{Bird1987}:
\begin{equation}
	\tau_R(\dot{\gamma}_{a,R})=\frac{\mathcal{M}}{2\pi R^3}\bigg[3+\frac{d\ln \mathcal{M}/2\pi R^3}{d\ln \dot{\gamma}_{a,R}} \bigg]
	\label{eq5}
\end{equation}
where $\mathcal{M}$ denotes the torque applied on the upper plate. The corrected flow curve, pictured as filled triangles (\textcolor{red}{$\blacktriangle$}) in Fig.~\ref{fig:8}(a) appears shifted towards lower stresses for $\dot \gamma \leq  7$~s$^{-1}$. Alternatively, single-point determination methods can also be used to correct for shear inhomogeneity to avoid additional noise associated with numerical calculation of the derivative in Eq.~(\ref{eq5}) \cite{Cross1987,Shaw2006}. These methods rely on calculating the apparent Newtonian viscosity at the rim $\eta_N=2\mathcal{M}H/\pi R^4 \Omega_{rot}$ which is then associated with a lower shear rate, equal to approximately three quarters of the rim shear rate. For Bingham plastic and power-law materials, the error associated with performing this single-point correction is typically less than 2\% \cite{Cross1987,Shaw2006}. Indeed, the result of the latter method is barely distinguishable from that of the derivative method  [(\textcolor{red}{$\blacktriangledown$}) in Fig.~\ref{fig:8}(a)]. Finally, correction for slip can be performed by conducting rheological measurements at different gaps \cite{Yoshimura1988, Clasen2012, Zahirovic2009}. If the flow curves obtained at various gaps superimpose, the material does not slip, whereas a gap-dependent rheology is the signature of wall slip. Assuming symmetric slip at both walls, the following kinematic relationship can be derived for a given imposed stress $\tau$ \cite{Yoshimura1988}: 
\begin{equation}
	\dot{\gamma}_a(\tau)=\dot{\gamma}_t(\tau)+\frac{2V_s(\tau)}{H}
	\label{eq7}
\end{equation}
where $\dot{\gamma}_a$ is the apparent shear rate, $\dot{\gamma}_t$ the true shear rate and $V_s$ the slip velocity. Experiments conducted at different gaps [see Fig.~S7 in the Supplemental Material] together with Eq.~(\ref{eq7}) allow us to compute the true flow curve of the carbon-black gel [($\blacksquare$) in Fig.~\ref{fig:8}(a)]. The corrections reveal that the gel exhibits a critical shear rate of about 1~s$^{-1}$ below which no homogeneous flow is possible (see additional discussion below about the formation of log-rolling flocs) and that, for shear stresses larger than 50~Pa, the shear rate was underestimated prior to the shear inhomogeneity correction. 

We now turn to the electric measurements, and discuss the corrections to the apparent conductivity defined in Eq.~(\ref{eq1}) which is observed to decrease for increasing shear rate [Fig.~\ref{fig:8}(b)]. First, following a calculation analogous to the one that leads to Eq.~(\ref{eq5}), a correction for shear inhomogeneity can be derived for the conductivity [see details in section~II A of the Supplemental Material] which reads for the apparent conductivity $\sigma_{a,R}$ at the rim:
\begin{equation}
	\sigma_{a,R}(\dot{\gamma}_{a,R})=\frac{I}{\phi_0}\frac{H}{2\pi R^2}\bigg[2+\frac{d\ln \big(\frac{I}{\phi_0} \frac{H}{2\pi R^2}\big)}{d\ln \dot{\gamma}_{a,R}} \bigg]
	\label{eq6}
\end{equation}
where $I$ denotes the measured current and $\phi_0$ the applied potential. Corrected data appears as simply shifted along the shear-rate axis [(\textcolor{red}{$\vartriangle$}) in Fig.~\ref{fig:8}(b)]. Moreover, an analogous calculation to the single-point determination method described in \cite{Cross1987,Shaw2006}, leads to a similar expression for the conductivity (see details in section~II B of the Supplemental Material). The apparent conductivity as defined in Eq.~(\ref{eq1}) is associated with a lower shear rate, equal to approximately two thirds of the shear rate at the rim. Here again, similarly to the rheological correction, the result of this  method is barely distinguishable from that of the derivative method [(\textcolor{red}{$\triangledown$}) in Fig.~\ref{fig:8}(b)]. Finally, the expression for the contact resistance shown in Eq.~(\ref{eq4}), which is the electrical equivalent to the correction for slip [Eq.~(\ref{eq7})] can be extended to a sample under flow [see section~II C in the Supplemental Material]. For a given stress at the rim $\tau_R$, one can thus write the resulting resistivity as:
\begin{equation}
	\frac{1}{\sigma_{a,R}(\tau_R)}=\frac{1}{\sigma_{b,R}(\tau_R)}+\frac{\rho_{c,R}(\tau_R)}{H}
	\label{eq8}
\end{equation}
where $\sigma_{a,R}$ denotes the measured apparent conductivity derived in Eq.~(\ref{eq7}) and $\sigma_{b,R}(\tau_R)$ is the bulk conductivity at the level of imposed stress $\tau_R$. Finally, $\rho_{c,R}$ is the specific contact resistance at the rim expressed in $\Omega$.cm$^2$ and defined as $\rho_{c,R}(\tau_R)=(\partial \mathcal{R}_c/\partial j)_R(\tau_R)$ where $\mathcal{R}_c$ is the contact resistance and $j$ is the current density. By introducing the specific contact resistance, Eq.~(\ref{eq8}) takes into account the potential variation of the contact resistance with the radius $r$ and hence its dependence on both the shear rate and shear stress. Ultimately, we find that the fully corrected conductivity curve shows the same trend as the raw data but is quantitatively shifted towards larger conductivities [($\square$) in Fig.~\ref{fig:8}(b)].

\begin{figure}
	\centering
	\resizebox{0.5\textwidth}{!}{\includegraphics{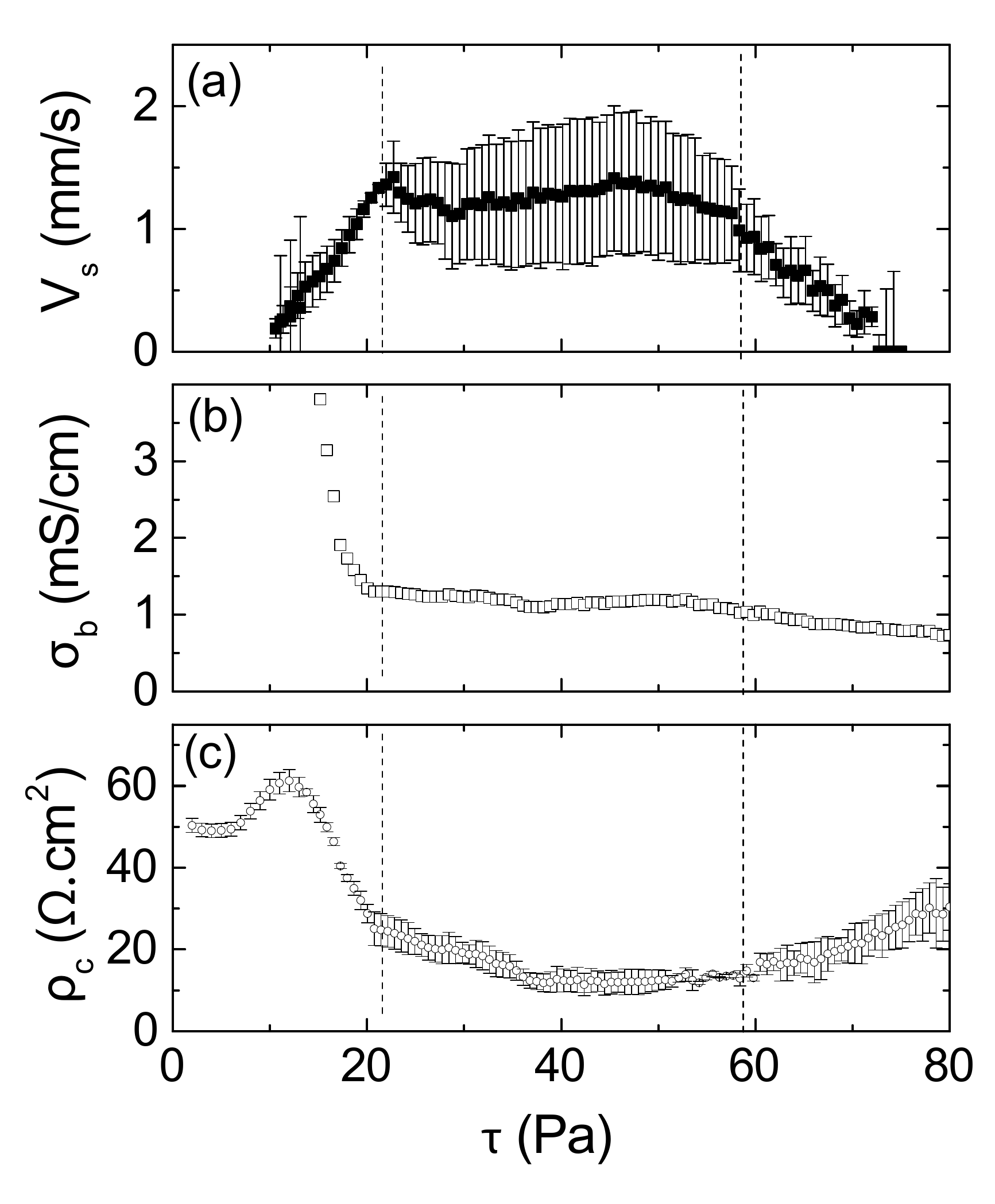}	}	
	\caption{Slip velocity $V_s$, bulk conductivity $\sigma_b$ and specific contact resistance $\rho_c$ obtained from the slip and contact resistance corrections performed on the steady state flow curves in a 8\%~wt. carbon black gel (VXC72R). The vertical dashed lines indicate the limits of the three regimes discussed in the text. From large to low stresses: fully fluidized regime for $\tau\geq55$ Pa, plug flow with slip at the walls for $55 \text{ Pa}\geq\tau \geq 20$ Pa and shear-induced structuration (formation of vorticity-aligned flocs) for $\tau\leq20$ Pa. (a) Slip velocity $V_s(\tau)$ (b) Bulk conductivity $\sigma_{b}(\tau)$ (c) Specific contact resistance $\rho_{c}(\tau)$.}
	\label{fig:9}       
\end{figure}

Experiments conducted at different gaps [see Fig.~S8 in the Supplemental Material] together with Eq.~(\ref{eq8}) allow us to compute the bulk conductivity of the carbon-black gel [($\square$) in Fig.~\ref{fig:8}(b)]. Let us now discuss the quantities extracted from both the mechanical and electrical corrections of the flow curve, i.e. the slip velocity $V_s$, the bulk conductivity $\sigma_b$ and the specific contact resistance $\rho_c$. These quantities are reported as a function of the applied stress $\tau$ in Fig.~\ref{fig:9} during the flow cessation of a 8\%~wt. CB gel (VXC72R). In consequence, the graphs in Fig.~\ref{fig:9} should be read from right to left. At large stresses, i.e. $\tau \geq 75$~Pa, the slip velocity is negligible, and the bulk resistance and specific contact resistance are constant, in agreement with a fully fluidized sample that is sheared homogeneously. As the stress is ramped down to $\tau\simeq 55$~Pa, the bulk conductivity slowly increases and the slip velocity becomes positive while the specific contact resistance decreases. These evolutions reflect the aggregation of the CB particles into larger flocs which better conduct the current and display an increasing amount of slip at the wall. Such aggregates experience more contact with the wall than individual CB particles, hence the decrease in $\rho_c$. From $\tau \simeq 55$~Pa to 20~Pa, both $\sigma_b$ and $V_s$ are roughly constant while $\rho_c$ slowly increases as the stress is ramped down. These results strongly suggest the appearance of thin lubrication layers at both walls, mainly composed of oil, and isolating a bulk percolated structure of CB particles, that is only weakly sheared. Moreover, assuming a slip layer of thickness $\delta$ and viscosity $\eta_s$ that is equal to the value for the matrix phase, the stress at the wall reads \cite{Barnes1995}: $\tau_w=\eta_s V_s/\delta$, where $V_s/\delta$ denotes the shear rate inside the slip layer. As both $V_s$ and $\delta$ remain constant while decreasing the stress, one should expect the thickness $\delta$ of the slip layer to progressively increase from $\delta \simeq 400$~nm to 1~$\mu$m. And indeed, the specific contact resistance increases concomitantly with the stress decrease. Finally, for $\tau \leq 20$ Pa, both the bulk conductivity and the contact resistance show a dramatic increase which is the signature of strong changes in the sample microstructure that we interpret as the formation of vorticity-aligned flocs \cite{Osuji2008b,Negi2009,Grenard2010}. 
Indeed, the formation of log rolling flocs has been reported in confined geometries for shear rates lower than a critical value $\dot{\gamma}_c$ that scales for VXC72R CB gels as $\dot \gamma _c=3650/H^{1.4}$ with $H$ given in $\mu$m, independently of the CB concentration \cite{Grenard2010}. For $H=750$~$\mu$m, one calculates $\dot \gamma_c \simeq 1.2$~s$^{-1}$ which, according to the true flow curve of the gel [Fig.~\ref{fig:8}(a)], corresponds to  $\tau_c \simeq 25$~Pa, in good agreement with the transition observed in Fig.~\ref{fig:9}, at $\tau \simeq 20$~Pa. Therefore at low shear rates, the rheo-electric measurements reflect the growth of a shear-induced structure which is otherwise hard to visualize at such large concentrations of CB \cite{Grenard2010}.

 The above analysis shows that conclusions drawn from rheo-electric measurements must be considered with care, and, we have highlighted the importance of corrections for shear inhomogeneity in parallel plate geometry, wall slip and contact resistance, which are often overlooked in the study of strongly conductive complex fluids. Application of these corrections also provides quantitative values for important quantities, namely the slip velocity, bulk conductivity and specific contact resistance which, once combined, provide a more complete picture of the flow profile in the conductive gel at steady state over a large range of stress values. \\

\section{Discussion}
\label{sec:4}

The rheo-electric apparatus described in section~\ref{sec:1} provides robust measurements for highly conductive materials at rest, during transient flows and under steady-state shear. 

Flow cessation experiments on carbon black gels reported in section~\ref{sec:3.1} have revealed that the sample shearing history tunes concomitantly the elastic and the electric properties of these attractive gels. Progressive stress decrease down to $\tau=0$~Pa leads to a more solid-like gel whose elastic modulus $G'_0$ varies as a power-law function of the conductivity $\sigma_0$ for increasing quenching speed: i.e. $G'_0 \propto \sigma_0^{\alpha}$, with $\alpha=1.65\pm 0.04$. Previous measurements on gels formulated with exact same carbon black powder from Cabot at various volume fractions $\phi$ have shown that: $G'_0 \propto (\phi-\phi_c)^{\nu}$, with $\nu=4.0\pm 0.1$, where the exponent $\nu$ is consistent with percolation theory \cite{Trappe2000}. Assuming that the electrical and elastic percolation occurs at the same volume fraction $\phi_c$, one can expect from the measurements reported here that: $\sigma_0 \propto (\phi-\phi_c)^t$, with $t=\alpha/\nu = 2.4\pm 0.1$ \cite{note1}. Interestingly, this value is quantitatively outside the range of values  predicted by percolation theories $1.5\leq t\leq 2$ \cite{Kirkpatrick1973,Straley1977,Clerc1990}, and larger than values recently reported for other carbon black gels: Ketjen Black ($t=1.9$) and Super C-45 ($t=1.7$) dispersed in an organic solvent \cite{Youssry2013}. Our measurements strongly suggest that the VXC72R carbon black gel exhibit an anomalous conductivity dependence. Such conclusions are strikingly similar to that obtained in Cabot carbon-black polymer composite, where CB particles are embedded in an insulating plastic \cite{Balberg1987}. Indeed, composites loaded with the Cabot carbon black show an exponent larger than 2, ($t=2.8\pm 0.2$ for CB particles embedded in polyvinylchloride) which has been interpreted as the consequence of a more compact network, due to the less tortuous shape of these particles \cite{Rubin1999,Balberg2002}. Our result shows here that such an exceptional behavior of the VXC72R also extends to carbon-black networks embedded in a non-aqueous solvent.  

We have also shown in section~\ref{sec:3.2} that the rheo-electric apparatus is well adapted to probe the evolution of the gel structure under weak deformations and provides much higher sensitivity for monitoring the evolution of the microstructure compared to other techniques such as velocimetry  \cite{Grenard2014}. Indeed, applied to creep experiments, rheoelectric measurements reveal that the power-law response of the carbon black gel during the primary creep regime is associated with an increase in the network connectivity and conductivity of the gel. Power-law responses to creep tests have been reported already for numerous other soft  materials such as microgels \cite{Divoux2011b}, polycrystalline surfactant hexagonal phases \cite{Bauer2006}, core-shell colloidal particles \cite{Siebenburger2012a} and protein gels \cite{Brenner2013,Leocmach2014}. However, the present measurements are among the first to illustrate the existence of simultaneous rearrangements of the sample in bulk. At the strand level, such rearrangements may likely correspond to the net creation of new bonds due to the presence of attractive van der Waals forces between the soot particles. Indeed, recent numerical simulations on model colloidal gels have revealed that imposed macroscopic deformation leads to the formation of new bonds even at low strain \cite{Colombo2014}, which may account for the increase of conductivity reported in Fig.~\ref{fig:7}. More simulations under applied shear stress conditions are needed to clarify the evolution of the sample network connectivity during primary creep.  

\begin{figure}[t!]
	\centering
	\resizebox{0.5\textwidth}{!}{\includegraphics{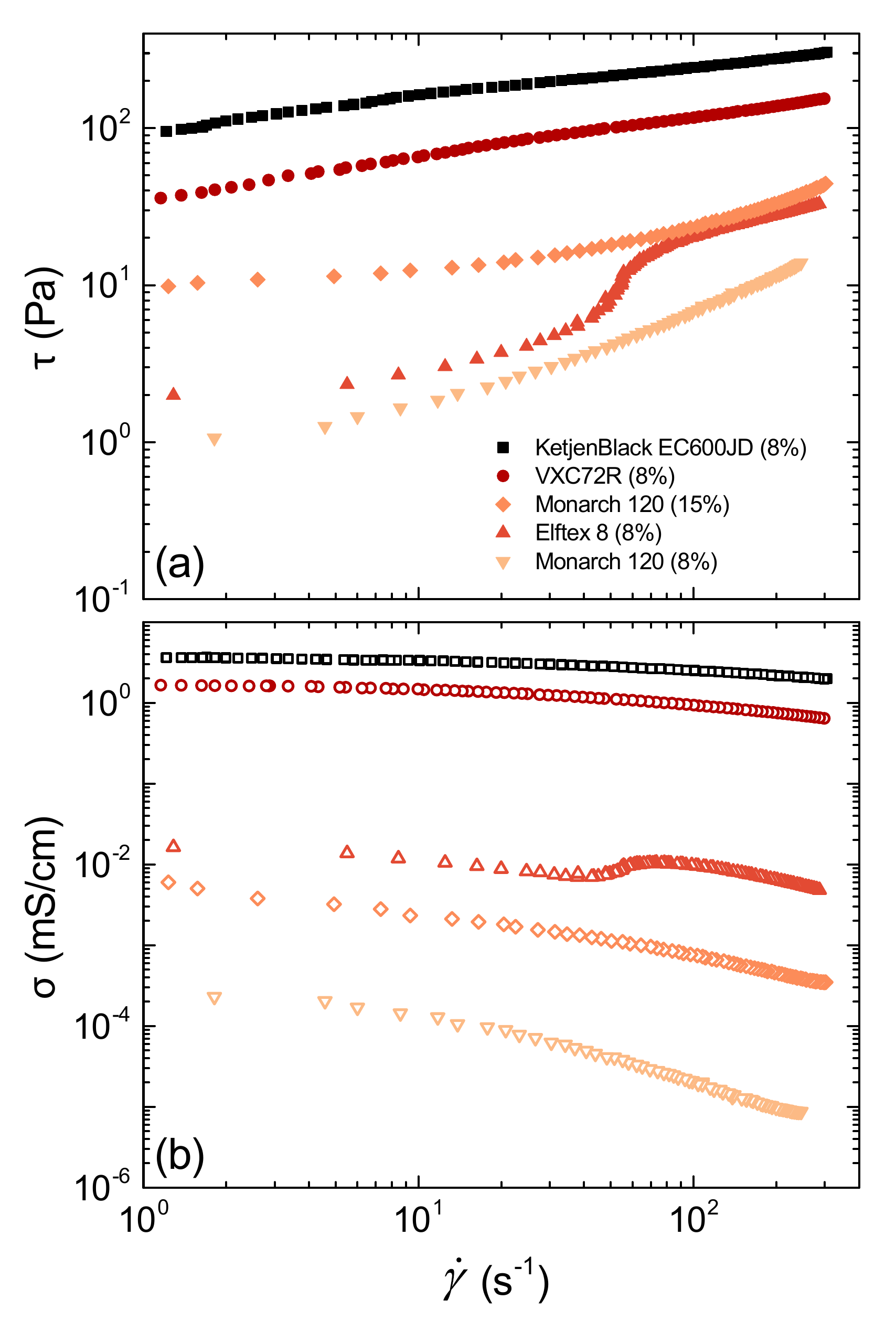}}	
	\caption{Sets of corrected (a) flow curves and (b) conductivity curves for carbon black gels made with carbon black particles of different grades: KetjenBlack EC600JD, VXC72R, Elftex 8 at weight concentration $c=8\%$ wt., and Monarch 120 at $c=8$\%~wt. and 15\%~wt.}
	\label{fig:10}       
\end{figure}

In steady shear, we have shown in section~\ref{sec:3.3} that performing corrections for shear inhomogeneity, wall slip and contact resistance are crucial to correctly determine the electrical and mechanical constitutive behavior of CB gels under flow. As a result, we can now safely compare the rheo-electric properties of CB gels made from different grades of carbon black. We examine 8\%~wt. CB gels made of respectively: VXC72R, Monarch 120, Elftex 8 (Cabot) and KetjenBlack EC600JD (Akzo Nobel), as well as a 15\%~wt. dispersion made of Monarch 120 (Cabot). Similarly to the experiments reported in Fig.~\ref{fig:4},  ramps of decreasing stress from $\tau=\tau_{max}$ down to $\tau=0$~Pa and composed of $N=100$ steps each of duration $\Delta t=15$~s are applied to each gel. The value of $\tau_{max}$, which ranges between 15~Pa and 300~Pa depending on the type of CB gels, is chosen such as the corresponding shear rate is 300~s$^{-1}$ for all samples (Fig.~\ref{fig:10}).  The rheo-electric properties of the different 8\%~wt. CB gels span over several orders of magnitude both in conductivity and stress. On the one hand, CB gels made with KetjenBlack EC600JD and VXC72R show a high conductivity larger than 1~mS/cm that is weakly sensitive to shear, which is indeed why these CB were initially designed. Such properties are intimately linked to their large specific surface area of about 200~m$^2/g$ for the VXC72R sample \cite{KyrlidisSawkaMoeserEtAl2011,Vulcan} and 1400~m$^2/g$ for the Ketjenblack EC600JD \cite{Fan2014, Duduta2011,KyrlidisSawkaMoeserEtAl2011}. 
On the other hand, CB gels made with Elftex 8 and Monarch 120 display a much lower conductivity that is also more strongly sensitive to shear. Correspondingly, these CB have a lower specific surface area of about 75~m$^2/g$ \cite{Elftex} and 29~m$^2/g$ \cite{Monarch} respectively. Such CB are typically used for color as pigments where CB conductive properties are less crucial. Finally, we note that the electrical properties of these CB dispersions are strongly related to the individual properties of the CB particles. In particular,  increasing the concentration in CB does not compensate for low conductivity values or shear sensitivity, as illustrated by the results for a gel made of 15\% wt. of Monarch 120. These measurements illustrate how the rheo-electric fixture can be used to compare, rank and select the desired properties of highly conductive complex fluids.\\

\section{Conclusion}
In this work, we have introduced a new rheo-electric apparatus designed for conventional stress-controlled rheometers. This apparatus can be used to accurately measure the conductivity  of strongly conductive materials under flow at the same time as the viscometric properties of the gel are measured. Benchmark tests on CB gels have allowed us to illustrate the high sensitivity of the apparatus even at low strains, and to determine the transient and steady-state behavior of CB gels by comparing the simultaneous electrical and mechanical response. Furthermore, by taking into account shear inhomogeneity, wall slip and the existence of a measurable contact resistance, we have shown that the correction methods first introduced to extract true bulk mechanical measurements from parallel plate rheometry data can be formally extended to analyze electrical data. The latter analysis should be used as a reference for future electrical characterization of conductive complex fluids such as conductive silver pastes, carbon nanotube nanocomposites, conductive inks, etc. More generally, the low friction rheo-electric apparatus we have designed for stress-controlled rheometers paves the way for a wider use of rheo-electrical measurements in the broad soft matter community. In particular, the present apparatus enables two-electrode electrochemical tests to be performed under flow, which is of primary importance for flow batteries, where the systematic characterization of cell performance under flow in a controlled environment is crucial for cell design and optimization. 

\begin{acknowledgments}
The authors thank X.-W. Chen, Y.-M. Chiang, E. Del~Gado, F. Fan and S. Manneville for insightful discussions, and TA instruments for technical advice. The authors acknowledge support from  the Joint Center for Energy Storage Research (JCESR), an Energy Innovation Hub funded by the U.S. Department of Energy and the Basic Energy Science (BES) program of the Office of Science (SC). Finally, the authors gratefully acknowledge support from the MIT-France seed fund and the CNRS through a PICS-USA scheme (\#36939). 
\end{acknowledgments}

\clearpage
\setcounter{figure}{0}

\begin{widetext}

\begin{center}
\Large{Supplemental Material to:\\
 ``Simultaneous rheo-electric measurements of strongly conductive complex fluids"}
\end{center}

\section*{I. Supplemental figures}

\begin{figure}[h]
	\centering
	\resizebox{0.6\textwidth}{!}{
		\includegraphics{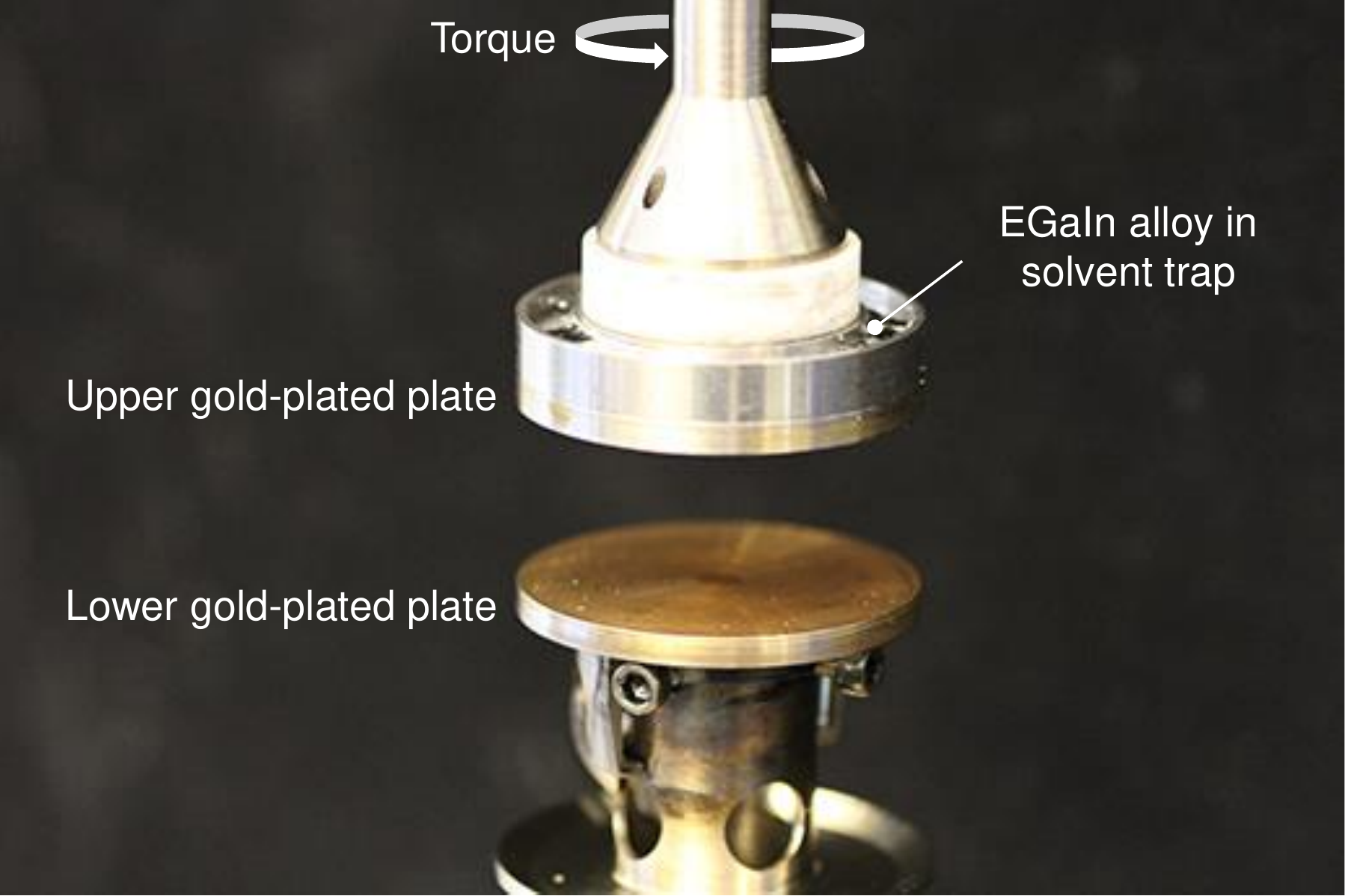}}
	\caption{Picture of the rheo-electric test fixture sketched in Fig.~1 in the main text. The geometry consists in parallel plates which are coated in gold to reduce the contact resistance. The solvent trap on top of the upper plate contains a liquid metal (EGaIn) which serves as a low-friction continuous electrical connection closing the electrical circuit, yet allowing the rotation of the upper plate with a minimum friction. The scale is fixed by the diameter of the plates ($2R=40$~mm). }
	\label{fig:s1}       
\end{figure}

\clearpage

\begin{figure}[h]
	\centering
	\resizebox{0.5\textwidth}{!}{\includegraphics{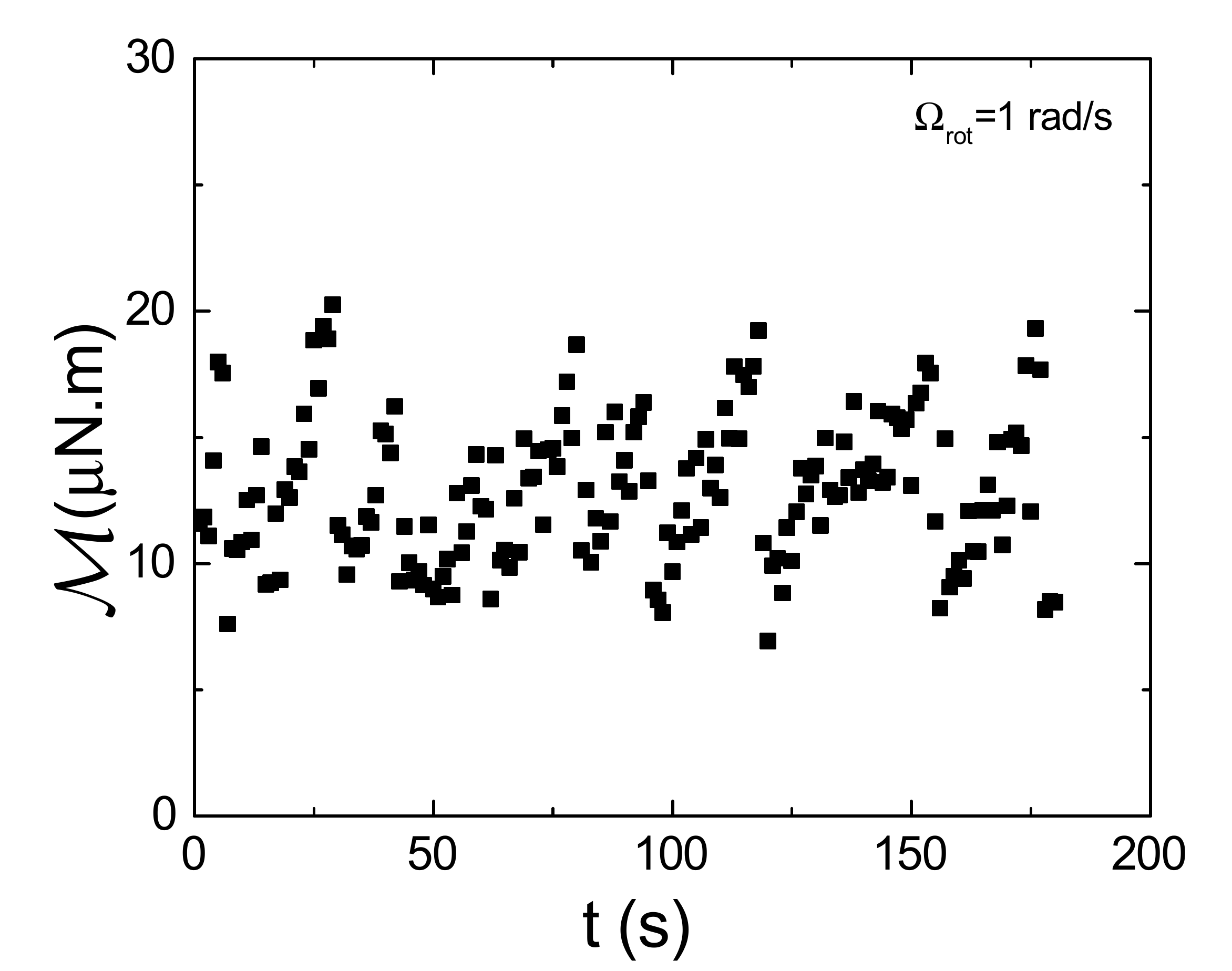}}
	\caption{Typical fluctuations of the frictional torque $\mathcal{M}$ from the EGaIn alloy as a function of time. Data recorded in steady state for an angular velocity $\Omega_{rot}=1$ rad/s of the upper plate, and with the liquid metal (EGaIn) ensuring the electrical contact. Such data are representative of the noise level associated with the rheological measurements when simultaneously conducting electrical measurements with the rheo-electric setup. For applied torques sufficiently large compared to $15\pm 5$~$\mu$N.m, i.e. for stresses here larger than about $1.2\pm 0.4$~Pa, such temporal fluctuations can be neglected. }
	\label{fig:s2}       
\end{figure}

\clearpage

\begin{figure*}[h]
	\centering
	\resizebox{1\textwidth}{!}{\includegraphics{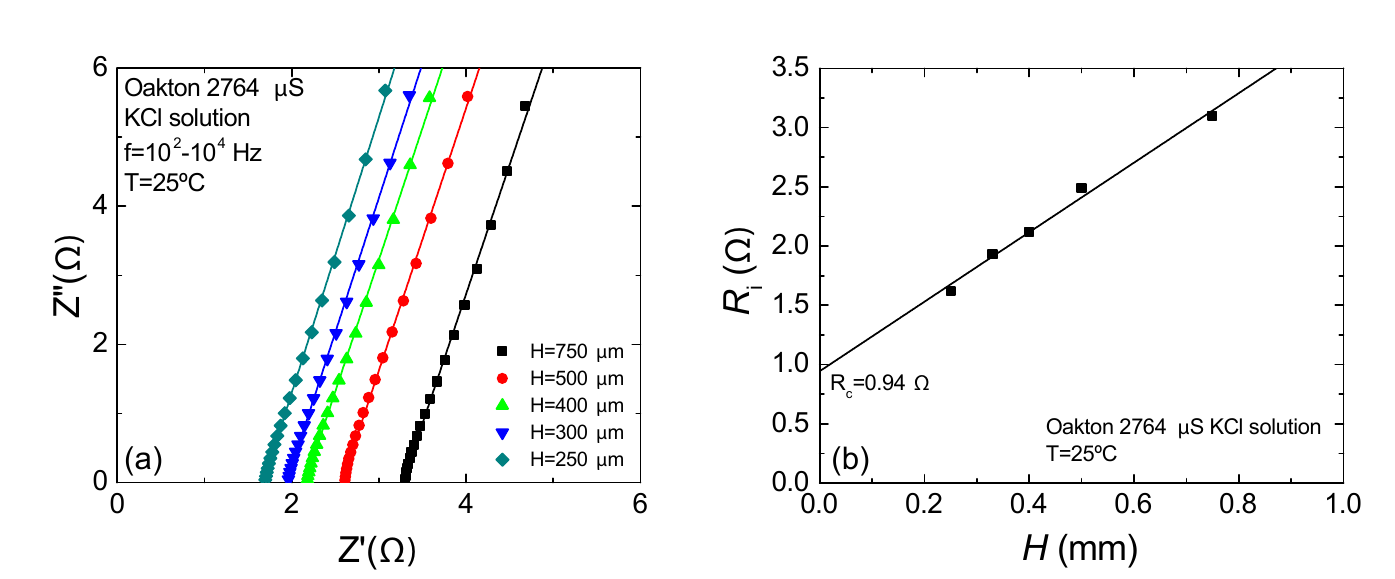}}
	\caption{(a) Nyquist plot of the complex impedance for an ionic conductor (Oakton 2764 $\mu$S/cm KCl solution) measured under static conditions for various gap heights $H=750, 500, 400, 330$ and 250~$\mu$m. This suspension is characterized by an ohmic resistor in parallel with an imperfect capacitor [$Z^*(\omega)=\mathcal{R}_i+1/(i\omega C_i)^{\alpha}$], represented by a constant phase element of phase $\alpha$ and which appears in the Nyquist plot as a line of slope $\tan(\alpha\pi/2)$ that intercepts with the horizontal axis at $Z'=\mathcal{R}_i$. Color lines pictured here correspond to the best linear fit of the data for each gap setting. (b) Ionic resistance $\mathcal{R}_i$ as determined in (a) vs the gap height $H$. The present data set shows that the sample resistance is divided into two contributions: a contact resistance $\mathcal{R}_{c}$ which corresponds to the intercept, and a bulk resistance associated with the slope; see Eq.~(4) in the main text. The black line is the best linear fit of the data and leads to: $\mathcal{R}_c=0.94$~$\Omega$ and $\sigma_b=(2.76\pm 0.01)$~mS/cm. }
	\label{fig:s3}       
\end{figure*}

\clearpage

\begin{figure*}[h]
	\centering
	\resizebox{1\textwidth}{!}{\includegraphics{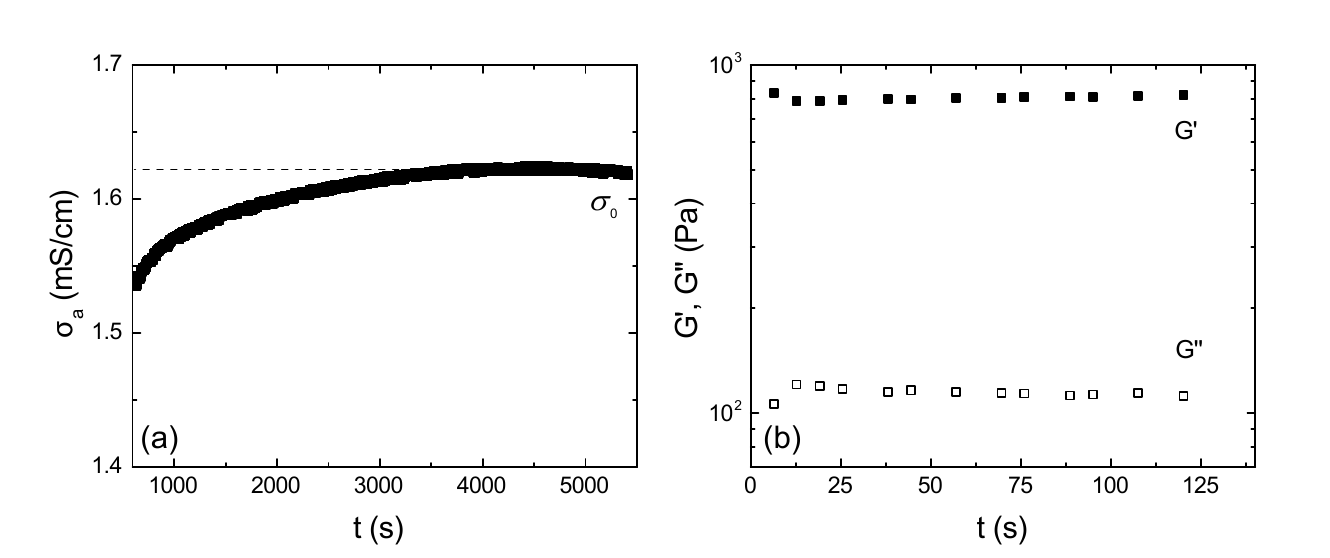}}
	\caption{(a) Temporal evolution of the apparent conductivity $\sigma_a$ in a 8\% wt. carbon black gel (VXC72R) during the 90~min recovery period that follows a decreasing ramp of stress, from $\tau=100$~Pa to $\tau=0$~Pa. The stress and conductivity data associated with the ramp are pictured in Fig.~4 in the main text (duration per point $\Delta t=6$~s). The conductivity reaches a steady-state value $\sigma_0=$1.62 mS/cm after about an hour. (b) Viscoelastic moduli $G'$ and $G''$ vs. time, measured with oscillations of small amplitude ($\gamma_0= 0.2$~\%) at $\omega= 6.3$~rad/s, after the 90~min recovery period pictured in (a). The viscoelastic moduli are constant with values $G'_0=805\pm 4$ ~Pa, and $G''_0=114\pm 6$~Pa. The values of $\sigma_0$ and $G'_0$ are reported in Fig.~5 in the main text, and further determined systematically for different  ramp durations $\Delta t$ in carbon black gels of various concentrations to assemble the data set reported in Fig.~5 in the main text.  }
	\label{fig:s4}       
\end{figure*}

\clearpage
\begin{figure}[h]
	\centering
	\resizebox{0.6\textwidth}{!}{\includegraphics{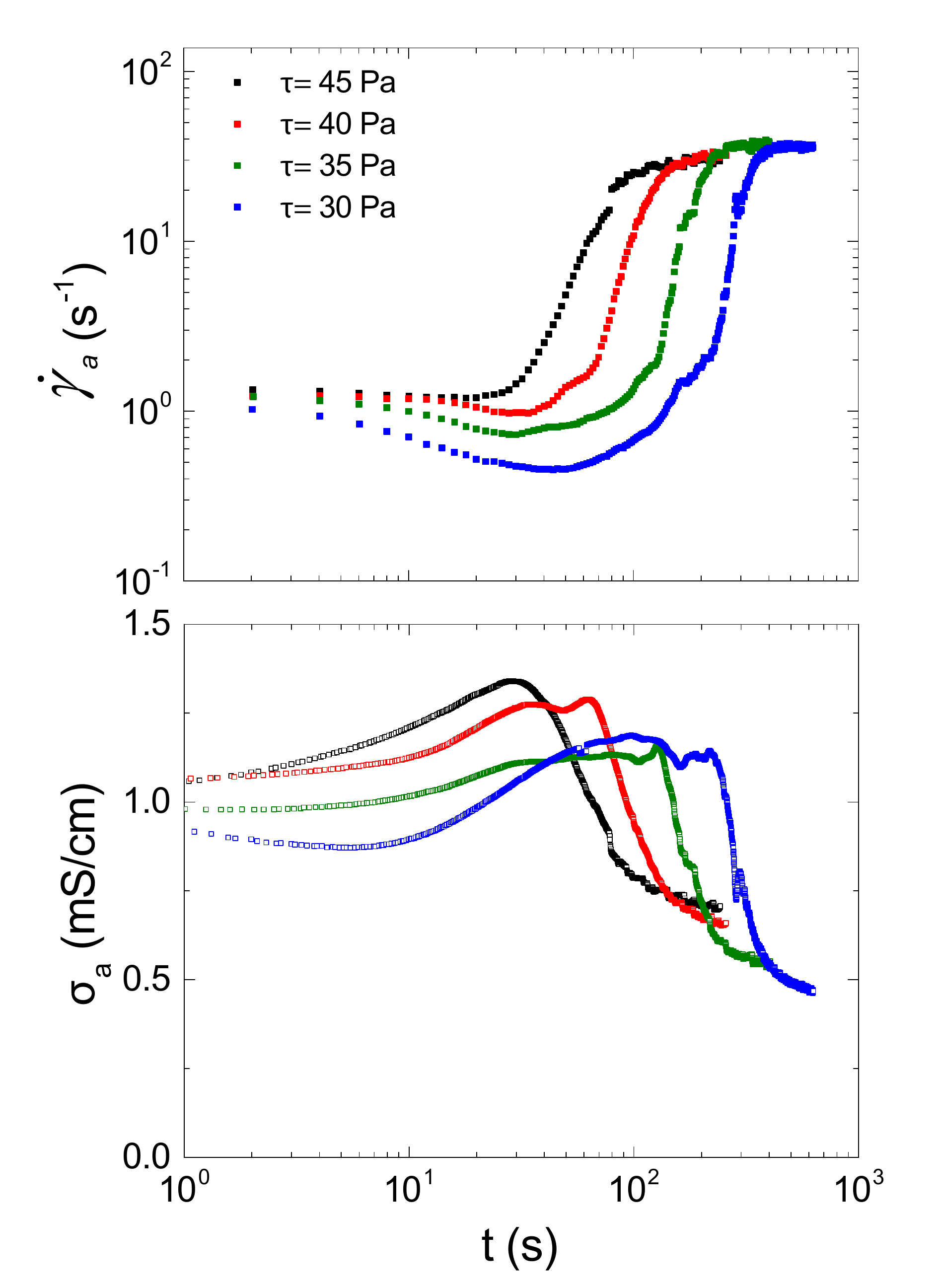}}
	\caption{Creep experiments in a 8\%~wt. carbon black gel (VXC72R) at various stresses $\tau=45, 40, 35$ and 30~Pa. (a) Shear-rate response $\dot \gamma_a(t)$ and (b) apparent conductivity response $\sigma_{a}(t)$. Before each experiment, the gel is presheared at $\dot \gamma=200$~s$^{-1}$ for 5~min to erase previous flow history, and then left to restructure for a waiting time $t_w=1200$~s at $\tau=0$~Pa.}
	\label{fig:s5}       
\end{figure}

\begin{figure}[h]
	\centering
	\resizebox{0.6\textwidth}{!}{\includegraphics{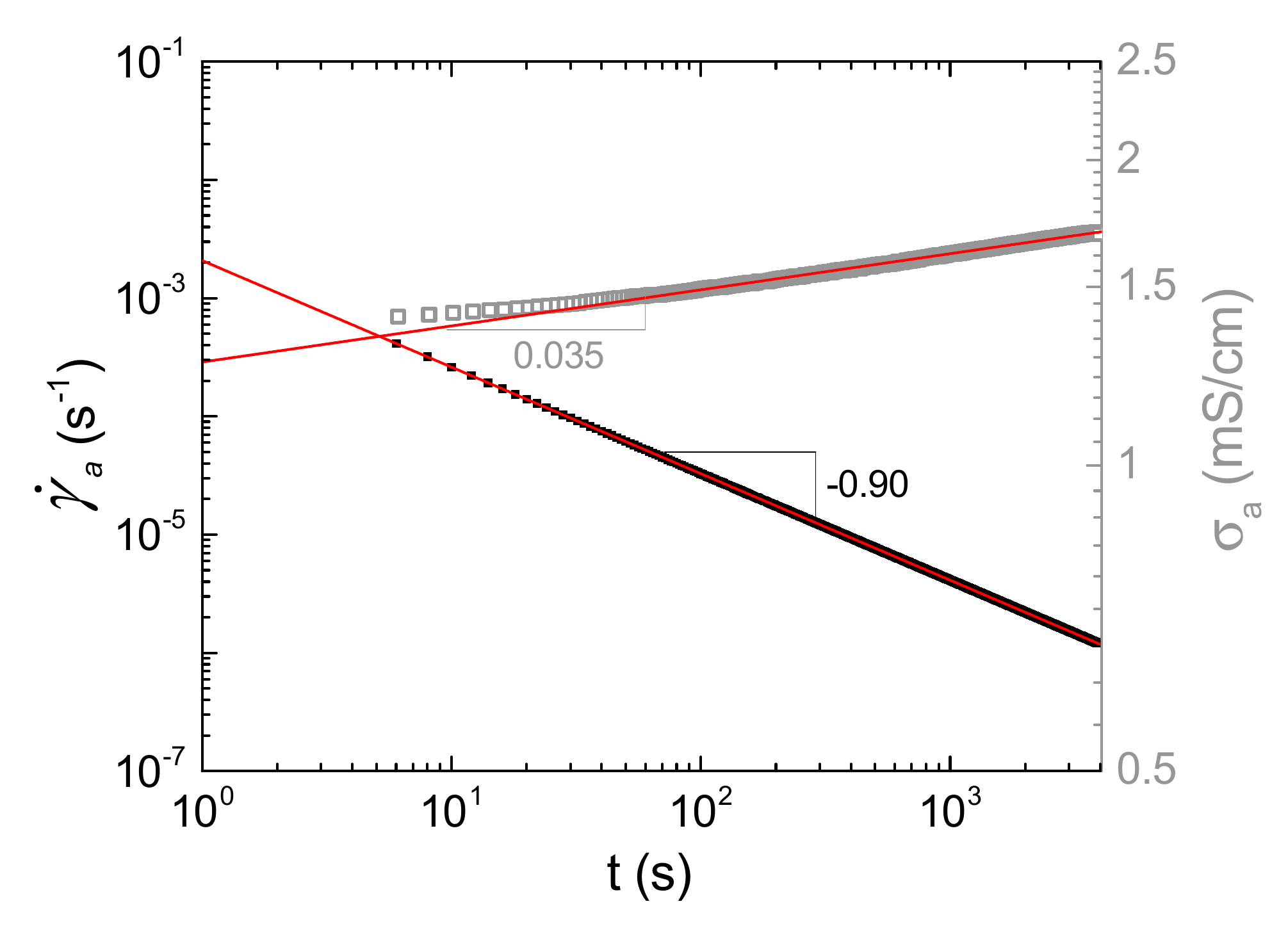}}
	\caption{Creep experiment in a 8\%~wt. carbon black gel (VXC72R) at $\tau=22$~Pa. Apparent shear-rate response $\dot \gamma_a(t)$ and apparent conductivity response $\sigma_{a}(t)$ on doubly-logarithmic scales to emphasize the power law characteristics of the primary creep regime. The same data is shown as that presented in Fig.~6(a) in the main text. The shear rate was calculated numerically from the strain measured by the rheometer to increase the precision. The red lines correspond to the best power-law fits of the data that leads to $\dot {\gamma}_a(t)=At^{\alpha}$ with $A=(2.3\pm 0.1)$ $10^{-3}$~$\text{s}^{-\alpha-1}$ and $\alpha=-0.90\pm 0.02$ and the $\sigma_a(t)=Bt^{\beta}$ with $B=1.26\pm 0.01$ mS$\text{.s}^{-\beta}$/cm and $\beta=0.035\pm 0.002$. We note that the power law fit of the conductivity is hardly distinguishable from a logarithmic fit $\sigma_a(t)=C\ln(t)+D$ for the data under study with $ C=0.0560 \pm 0.0005$ mS/cm and $D=1.23 \pm 0.01$ mS/cm.}  
\label{fig:s6}       
\end{figure}

\clearpage
\begin{figure*}[h]
	\centering
	\resizebox{1\textwidth}{!}{\includegraphics{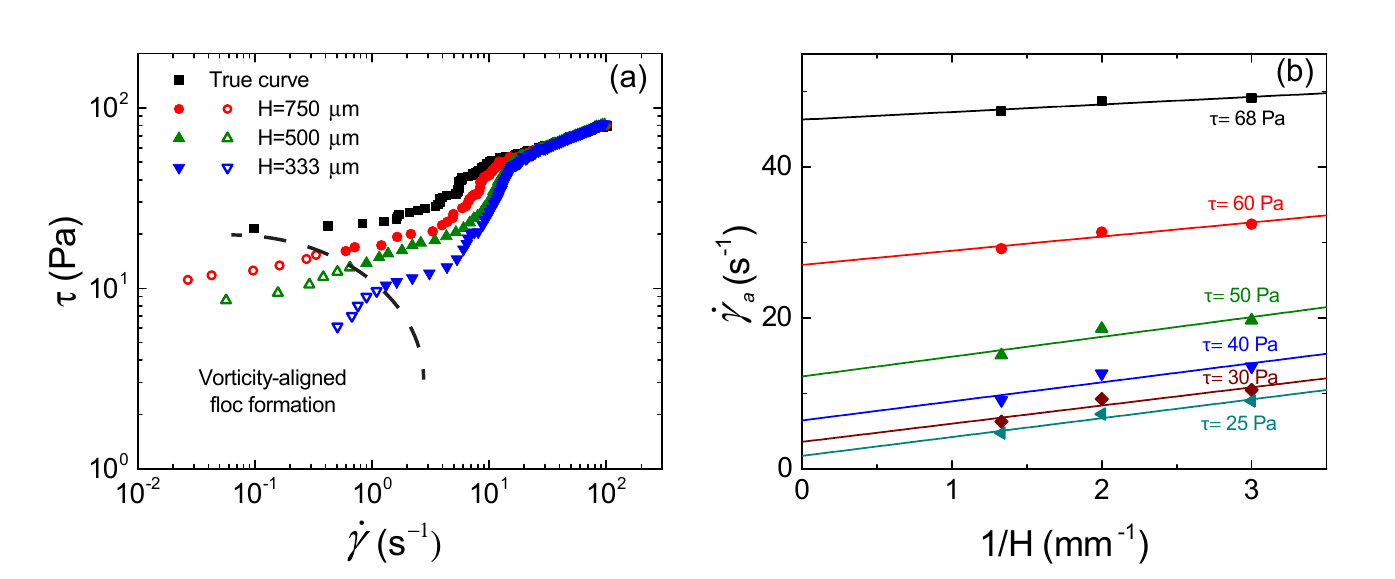}}
	\caption{(a) Flow curves of a 8\%~wt. carbon black gel (VXC72R) measured for various gap heights $H=750, 500$ and 330~$\mu$m together with the true flow curve corrected for slip. The open symbols indicate the region associated with the formation of vorticity aligned flocs. The boundary for the onset of this regime is given by  $\dot \gamma _c=3650/H^{1.4}$ with $H$ given in $\mu$m, independently of the CB concentration \cite{Grenard2010p}. (b) Plot of the measured values of apparent shear rate $\dot{\gamma}_a$ vs. $1/H$ for various imposed stresses $\tau=68, 60, 50, 40, 30$ and 25~Pa. Colored lines correspond to the best linear fits of the data and allow us to extract the true shear rate $\dot{\gamma}_t$ and the slip velocity $V_s$ for each imposed stress $\tau$ using Eq.~(6) in the main text. }
	\label{fig:s7}       
\end{figure*}

\begin{figure*}[h]
	\centering
	\resizebox{1\textwidth}{!}{\includegraphics{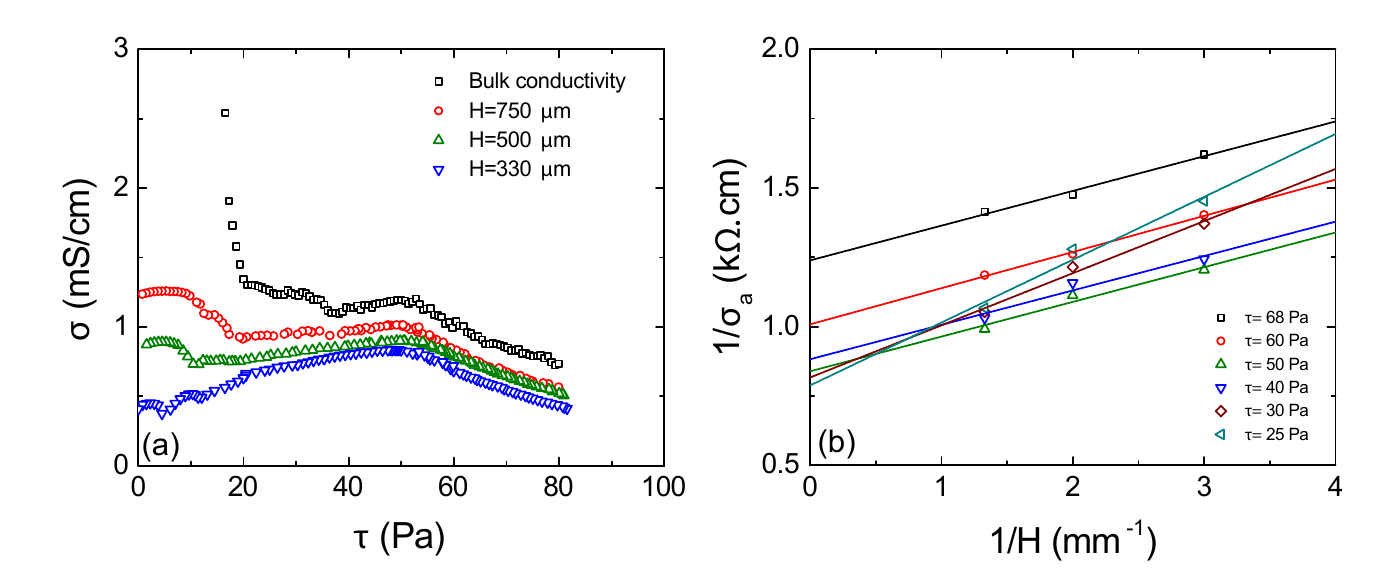}}
	\caption{(a) Conductivity curves of a 8\%~wt. carbon black gel (VXC72R) measured for various gap heights $H=750, 500$ and $330$~$\mu$m together with the bulk conductivity curve. (b) Inverse of the apparent resistivity $1/\sigma_a$ vs $1/H$ for various stresses $\tau=68, 60, 50, 40, 30$ and 25~Pa. Color lines correspond to the best linear fits of the data and allow to extract the bulk resistivity  $1/\sigma_b$ and the specific contact resistance $\rho_c$ for each stress $\tau$ using Eq.~(8) in the main text. }
	\label{fig:s8}       
\end{figure*}

\clearpage
\section*{II. Derivation of electrical corrections in the parallel plate}
\subsection{Shear inhomogeneity: derivative method}
In a parallel plate geometry, the shear rate $\dot{\gamma}$ is not homogeneous across the gap, but depends on the radial position $r$ as $\dot{\gamma}(r)=r\Omega_{rot}/H$, where $\Omega_{rot}$ is the angular velocity of the upper plate, and $H$ is the sample gap. Shear inhomogeneity can be corrected by means of the shear rate at the rim $\dot{\gamma}_{R}=R\Omega_{rot}/H$ and computing the stress at the rim $\tau_R$ using the following expression \cite{Bird1987p}:
\begin{equation}
	\tau_R(\dot{\gamma}_{R})=\frac{\mathcal{M}}{2\pi R^3}\bigg[3+\frac{d\ln \mathcal{M}/2\pi R^3}{d\ln \dot{\gamma}_{R}} \bigg]
	\label{eq1}
\end{equation}

where $\mathcal{M}$ denotes the torque applied on the upper plate.
A similar expression can be derived for electrical measurements. Indeed, the total current $I$ that is measured is given by the following expression:
\begin{equation}
	I= \frac{\phi_0}{H} 2\pi \int_{0}^{R}\sigma_a(r)rdr
	\label{eq2} 
\end{equation} 
where $\phi_0$ is the applied potential and $\sigma_a$ is the apparent conductivity. 
By changing the integration variable from $r$ to $\dot{\gamma}$, we obtain 
\begin{equation}
I= \frac{\phi_0 }{H} \frac{2\pi R^2}{{\dot{\gamma}_R}^2}\int_{0}^{\dot{\gamma}_R}\sigma_a(\dot{\gamma})\dot{\gamma} d\dot{\gamma} \,\, .
\label{eq3}
\end{equation}
Differentiating this result with respect to $\dot{\gamma}_R$, while using the Leibniz formula leads to  
\begin{equation}
\frac{dI}{d\dot{\gamma}_R}= \frac{-2I}{\dot{\gamma}_R}+\frac{\phi_0 }{H} \frac{2\pi R^2}{{\dot{\gamma}_R}} \sigma_{a,R}(\dot{\gamma}_R)  \,\, ,
\label{eq4}
\end{equation} 
that can be rewritten into: 
\begin{equation}
	\sigma_{a,R}(\dot{\gamma}_{R})=\bigg(\frac{H}{2\pi R^2}\frac{I}{\phi_0}\bigg)\bigg[2+\frac{d\ln \big(\frac{H}{2\pi R^2}\frac{I}{\phi_0}\big)}{d\ln \dot{\gamma}_{R}} \bigg]
	\label{eq5}
\end{equation}
The latter expression relates the apparent conductivity at the rim of the plate $\sigma_{a,R}$ to the macroscopic current $I$ measured at a shear rate $\dot{\gamma}_R=R\Omega_{rot}/H$. Eq.~(\ref{eq5}) allows us to correct for the radial shear inhomogeneity in the parallel plate geometry. 

\subsection{Shear inhomogeneity: Single-point method}
The single-point correction method relies on the Gaussian quadrature of moments equation which takes the following form \cite{Shaw2006p}:
\begin{equation}
\int_{0}^{1}f(x)x^k dx= \sum_{i=1}^{n}w_i f(x_i)
\label{eq6}
\end{equation}
where $k$ and $n$ denote integer larger than 1 and the pairs \{$w_i$,$x_i$\} are given in \cite{abramowitz1966p}
In the parallel plate geometry, the torque $\mathcal{M}$ can be expressed as an integral:
\begin{equation}
\mathcal{M}=2\pi\int_{0}^{R}\tau(r)r^2dr
\label{eq7}
\end{equation}
The integral can be non-dimensionalized by replacing the integration variable $r$ by $x=\dot{\gamma}/\dot{\gamma}_R=r/R$. We obtain 
\begin{equation}
\mathcal{M}=2\pi R^3\int_{0}^{1}\tau(x)x^2dx
\label{eq8}
\end{equation}
By using Eq.~(\ref{eq6}) with $n=1$ and $k=2$, we compute \{$w_i$,$x_i$\}= \{$1/3$,$3/4$\}, which leads to:
\begin{equation}
\mathcal{M}\simeq 2\pi R^3 \frac{\tau(3/4)}{3}
\label{eq9}
\end{equation}
otherwise written as:
\begin{equation}
\boxed{\tau \left(\frac{3\dot{\gamma}_R}{4}\right)\simeq\frac{3\mathcal{M}}{ 2\pi R^3} 
\label{eq10}}
\end{equation}

Eq.~(\ref{eq10}) links the stress at $r=3R/4$ to the torque $\mathcal{M}$. The error associated with Eq.~(\ref{eq10}) is generally less than 5\% for Bingham plastic and Newtonian fluids \cite{Shaw2006p}. Moreover, an expression similar to Eq.~(\ref{eq8}) can be derived for electrical measurements. Indeed, we can express the macroscopic current as follows: 
\begin{equation}
I= \frac{\phi_0 2\pi R^2}{H} \int_{0}^{1}\sigma_a(x)x dx
\label{eq11}
\end{equation}
Using again Eq.~(\ref{eq6}) with $n=1$ and $k=1$, we find \{$w_i$,$x_i$\}= \{$1/2$,$2/3$\} and thus that
\begin{equation}
I \simeq \frac{\phi_0 \pi R^2}{H} \sigma_a(2/3)
\label{eq12}
\end{equation}
which leads to the following expression:
\begin{equation}
\boxed{\sigma_a(\frac{2\dot{\gamma}_R}{3})\simeq\frac{I H}{\phi_0 \pi R^2} 
	\label{eq13}}
\end{equation}
which relates the apparent conductivity at $r=2R/3$ to the macroscopic current $I$ at a rotation rate $\Omega_{rot}$ and a gap $H$. The static apparent conductivity given by the value of $(I H/\phi_0 \pi R^2)$ is thus associated with the shear rate $2\dot{\gamma}_R/3=2\Omega_{rot}R/3H$.

\subsection{Contact resistance correction}

Viscometric corrections for wall slip can be performed by conducting rheological measurements at different gaps \cite{Yoshimura1988p, Clasen2012p, Zahirovic2009p}. If the flow curves obtained at various gaps superimpose, the material does not slip, whereas a gap-dependent rheology is the signature of wall slip. Assuming symmetric slip at both walls, the following kinematic relationship can be derived for a given imposed stress $\tau$ \cite{Yoshimura1988p}: 
\begin{equation}
	\dot{\gamma}_a(\tau)=\dot{\gamma}_t(\tau)+\frac{2V_s(\tau)}{H}
	\label{eq14}
\end{equation}
where $\dot{\gamma}_a$ is the apparent shear rate, $\dot{\gamma}_t$ the true shear rate and $V_s$ denotes the slip velocity.
In the case of a static measurement of conductivity, the apparent conductivity $\sigma_a$ measured experimentally is the sum of a bulk contribution $\sigma_b$ and a contact resistance $\mathcal{R}_c$ originating from the two interfaces:
\begin{equation}
	\mathcal{R}_{measured}\equiv\frac{\phi_0}{I}=\frac{H}{\sigma_{a}\pi R^2}=\mathcal{R}_{c}+\frac{H}{\sigma_{b}\pi R^2}
	\label{eq15}
\end{equation}
To extend the latter correction to a sample flowing in the parallel plate geometry, we consider an annulus of width ${\rm d}r$ at a radius $r$ and define the specific contact resistance at radius $r$ as $\rho_{c}(r)=(\partial \mathcal{R}_c(r)/\partial j)$,  where $\mathcal{R}_c$ is the contact resistance at a radius $r$ and $j=I/\pi R^2$ is the current density. In this annular ring, the incremental conductance can be written as 
\begin{equation}
d\mathcal{G}(r)=\frac{2\pi r {\rm d}r}{\rho_{c}(r)}+\frac{\sigma_{b}(r) 2\pi r{\rm d}r}{H}
\label{eq16}
\end{equation}
and the current as
 \begin{equation}
 dI(r)=\phi_0 d\mathcal{G}(r)=\frac{\phi_0 r {\rm d}r}{\frac{\rho_{c}(r)}{2\pi }+\frac{H}{\sigma_{b}(r) 2\pi }}
 \label{eq17}
 \end{equation}
 Summing the above expressions, the resistance of each component adds in parallel and the macroscopic current measured reads 
\begin{equation}
I=\int_{0}^{R} \frac{2\pi \phi_0 r {\rm d}r}{\rho_{c}(r)+\frac{H}{\sigma_{b}(r)}}
\label{eq18}
\end{equation}
Furthermore, following the same derivation as for the mechanical correction [see Eq.~(\ref{eq2}), (\ref{eq3}) and (\ref{eq4})] we obtain  
\begin{equation}
\sigma_{a,R}=\frac{H}{2\pi R^2}\frac{I}{\phi_0}\bigg[2+\frac{d\ln \big(\frac{H}{2\pi R^2}\frac{I}{\phi_0}\big)}{d\ln \dot{\gamma}_{R}} \bigg]=\frac{1}{\frac{\rho_{c,R}}{H}+\frac{1}{\sigma_{b,R}}}
\label{eq19}
\end{equation}
In summary, by analogy to the slip correction in Eq.~(\ref{eq14}), for a given imposed stress at the rim $\tau_R$ as defined by Eq.~(\ref{eq1}), we can decouple the bulk contribution from the surface contribution to the conductivity as follows:
\begin{equation}
\boxed{\frac{1}{\sigma_{a,R}(\tau_R)}= \frac{\rho_{c,R}(\tau_R)}{H}+\frac{1}{\sigma_{b,R}(\tau_R)}}
\label{eq20}
\end{equation}

Alternatively, if we use single-point determination methods to compute the apparent conductivity, we can derive the following relation:
\begin{equation}
\boxed{\frac{1}{\sigma_{a}[\tau(\frac{2\dot{\gamma}_R}{3})]}= \frac{\rho_{c}[\tau(\frac{2\dot{\gamma}_R}{3})]}{H}+\frac{1}{\sigma_{b}[\tau(\frac{2\dot{\gamma}_R}{3})]}}
\label{eq21}
\end{equation}
We note that the use of Eq.~(\ref{eq21}) is only useful if the full flow curve is known (e.g. from a stress sweep) as information about the quantity $\tau(\frac{2\dot{\gamma}_R}{3})$ at a shear rate $\dot{\gamma}=2\dot{\gamma}_R/3$ is unknown \textit{a priori} since $\tau(\frac{3\dot{\gamma}_R}{4})$ is the quantity obtained from the single-point determination method applied to the rheological data.

\end{widetext}


\begin{thebibliography}{90}
\bibitem [{not()}]{note1}%
  \BibitemOpen
  \href@noop {} {}\bibinfo {note} {Note: Here, the variable $t$ denotes the
  critical exponent associated with the electrical percolation and not
  time.}\BibitemShut {Stop}%
\end{thebibliography}
\end{document}